\documentclass[sigconf,natbib=true, screen=true]{acmart}

\usepackage{graphicx}
\usepackage{acronym}
\usepackage{xcolor}
\usepackage{bbm}
\usepackage{mleftright}
\usepackage{subcaption}
\usepackage{listings}
\usepackage[inline]{enumitem}
\usepackage{numprint}
\usepackage{dsfont}

\usepackage{tikz}
\usetikzlibrary{positioning,shapes,arrows,arrows.meta,calc}
\definecolor{gRed}{HTML}{c5221f}
\definecolor{gYellow}{HTML}{f29900}
\definecolor{gGreen}{HTML}{188038}
\definecolor{gBlue}{HTML}{1967d2}
\definecolor{gGrey}{HTML}{9aa0a6}
\definecolor{g900grey}{HTML}{202124}
\definecolor{g900blue}{HTML}{174EA6}
\definecolor{g900red}{HTML}{A50E0E}
\definecolor{g900yellow}{HTML}{E37400}
\definecolor{g900green}{HTML}{0D652D}
\definecolor{g700grey}{HTML}{5F6368}
\definecolor{g700blue}{HTML}{1967d2}
\definecolor{g700red}{HTML}{c5221f}
\definecolor{g700yellow}{HTML}{f29900}
\definecolor{g700green}{HTML}{188038}
\definecolor{g500grey}{HTML}{9AA0A6}
\definecolor{g500blue}{HTML}{4285F4}
\definecolor{g500red}{HTML}{EA4335}
\definecolor{g500yellow}{HTML}{FBBC04}
\definecolor{g500green}{HTML}{34A853}
\definecolor{g400blue}{HTML}{669DF6}
\definecolor{g300green}{HTML}{81C995}
\definecolor{g300grey}{HTML}{DADCE0}
\definecolor{g300blue}{HTML}{8AB4F8}
\definecolor{g200green}{HTML}{A8DAB5}
\definecolor{g200blue}{HTML}{AECBFA}
\definecolor{g100grey}{HTML}{F1F3F4}
\definecolor{g100green}{HTML}{CEEAD6}
\definecolor{g100blue}{HTML}{D2E3FC}
\definecolor{g050green}{HTML}{E6f4EA}
\definecolor{g050blue}{HTML}{E8F0FE}
\definecolor{g100red}{HTML}{FAD2CF}
\definecolor{g100yellow}{HTML}{FEEFC3}
\definecolor{g100green}{HTML}{CEEAD6}

\tikzset{
    neuralnetwork/.pic = {
        \node[draw, draw=black, minimum height=0.4cm, minimum width=0.25cm, text width=0.2cm, align=center, fill=black!5, rounded corners=4pt] (llm_box) at (0, 0) {};
        
        \path[draw=black!50] (-0.5, 0.150) -- (-0.30,  0.150);
        \path[draw=black!50] (-0.5,-0.150) -- (-0.30, -0.150);
        
        \path[draw=black!50] (-0.30, 0.150) -- (0,  0.30);
        \path[draw=black!50] (-0.30, 0.150) -- (0,  0   );
        \path[draw=black!50] (-0.30, 0.150) -- (0, -0.30);
        \path[draw=black!50] (-0.30,-0.150) -- (0,  0.30);
        \path[draw=black!50] (-0.30,-0.150) -- (0,  0   );
        \path[draw=black!50] (-0.30,-0.150) -- (0, -0.30);
        
        \path[draw=black!50] (0,-0.30) -- (0, 0   );
        \path[draw=black!50] (0, 0   ) -- (0, 0.30);
        
        \path[draw=black!50] (0, 0.30) -- (0.30,  0.150);
        \path[draw=black!50] (0, 0.30) -- (0.30, -0.150);
        \path[draw=black!50] (0, 0   ) -- (0.30,  0.150);
        \path[draw=black!50] (0, 0   ) -- (0.30, -0.150);
        \path[draw=black!50] (0,-0.30) -- (0.30,  0.150);
        \path[draw=black!50] (0,-0.30) -- (0.30, -0.150);
        
        \path[draw=black!50] (0.30, 0.150) -- (0.5, 0.150);
        \path[draw=black!50] (0.30,-0.150) -- (0.5,-0.150);
        
        \draw[draw=black!50,fill=black!25](-0.30,  0.150) circle (2.25 pt);
        \draw[draw=black!50,fill=black!25](-0.30, -0.150) circle (2.25 pt);
        \draw[draw=black!50,fill=black!25]( 0   ,  0.30 ) circle (2.25 pt);
        \draw[draw=black!50,fill=black!25]( 0   ,  0    ) circle (2.25 pt);
        \draw[draw=black!50,fill=black!25]( 0   , -0.30 ) circle (2.25 pt);
        \draw[draw=black!50,fill=black!25]( 0.30,  0.150) circle (2.25 pt);
        \draw[draw=black!50,fill=black!25]( 0.30, -0.150) circle (2.25 pt);
    }
}

\acrodef{LLM}{large language model}
\acrodef{IR}{information retrieval}
\acrodef{DCG}{discounted cumulative gain}
\acrodef{PRP}{pairwise relevance prompting}
\acrodef{LTR}{learning-to-rank}
\acrodef{nDCG}{normalized DCG}

\AtBeginDocument{%
  }

\copyrightyear{2025}
\acmYear{2025}
\setcopyright{cc}
\setcctype{by}
\acmConference[SIGIR '25]{Proceedings of the 48th International ACM SIGIR Conference on Research and Development in Information Retrieval}{July 13--18, 2025}{Padua, Italy}
\acmBooktitle{Proceedings of the 48th International ACM SIGIR Conference on Research and Development in Information Retrieval (SIGIR '25), July 13--18, 2025, Padua, Italy}\acmDOI{10.1145/3726302.3730051}
\acmISBN{979-8-4007-1592-1/2025/07}

\makeatletter
\gdef\@copyrightpermission{
\begin{minipage}{0.3\columnwidth}
\href{https://creativecommons.org/licenses/by/4.0/}{\includegraphics[width=0.90\textwidth]{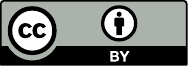}}
\end{minipage}\hfill
\begin{minipage}{0.7\columnwidth}
\href{https://creativecommons.org/licenses/by/4.0/}{This work is licensed under a Creative Commons
Attribution International 4.0 License.}
\end{minipage}
\vspace{5pt}
}
\makeatother

\ccsdesc[500]{Information systems~Search engine architectures and scalability}
\ccsdesc[500]{Information systems~Retrieval models and ranking}
\ccsdesc[500]{Information systems~Learning to rank}

\keywords{Learning to Rank, Large Language Models, Ranking Distillation}

\begin{document}

\allowdisplaybreaks

\title{Optimizing Compound Retrieval Systems}

\settopmatter{authorsperrow=4, printfolios=true}

\author{Harrie Oosterhuis}
\affiliation{
    \institution{Radboud University}
    \city{Nijmegen}
    \country{NL}
}
\email{harrie.oosterhuis@ru.nl}
\authornote{Work done while Harrie Oosterhuis was working at Google DeepMind.}

\author{Rolf Jagerman}
\affiliation{
    \institution{Google DeepMind}
    \city{New York}
    \country{US}
}
\email{jagerman@google.com}

\author{Zhen Qin}
\affiliation{
    \institution{Google DeepMind}
    \city{New York}
    \country{US}
}
\email{zhenqin@google.com}

\author{Xuanhui Wang}
\affiliation{
    \institution{Google DeepMind}
    \city{Mountain View}
    \country{US}
}
\email{xuanhui@google.com}

\begin{abstract}
Modern retrieval systems do not rely on a single ranking model to construct their rankings.
Instead, they generally take a cascading approach where a sequence of ranking models are applied in multiple re-ranking stages.
Thereby, they balance the quality of the top-$K$ ranking with computational costs by limiting the number of documents each model re-ranks.
However, the cascading approach is not the only way models can interact to form a retrieval system.

We propose the concept of \emph{compound retrieval systems} as a broader class of retrieval systems that apply multiple prediction models.
This encapsulates cascading models but also allows other types of interactions than top-$K$ re-ranking.
In particular, we enable interactions with \acp{LLM} which can provide relative relevance comparisons. %
We focus on the optimization of compound retrieval system design which uniquely involves learning where to apply the component models and how to aggregate their predictions into a final ranking.
This work shows how our compound approach can combine the classic BM25 retrieval model with state-of-the-art (pairwise) LLM relevance predictions, while optimizing a given ranking metric and efficiency target.
Our experimental results show optimized compound retrieval systems provide better trade-offs between effectiveness and efficiency than cascading approaches, even when applied in a self-supervised manner.

With the introduction of compound retrieval systems, we hope to inspire the information retrieval field to more out-of-the-box thinking on how prediction models can interact to form rankings.

\end{abstract}

\maketitle

\acresetall

\section{Introduction}

Modern retrieval systems consist of many components that are applied together to construct rankings of documents \citep{croft2010search}.
Thereby, they can combine different prediction models that provide different ranking utility at different costs, such that their service is both responsive and provides high quality search results~\citep{wang2011cascade, liu2017cascade}.
Accordingly, it appears very important for the \ac{IR} field to consider the following question:
\vspace{-0.1\baselineskip}
\begin{enumerate}[leftmargin=0em]
\item[] \emph{How should retrieval systems utilize multiple prediction models to form rankings in the most effective and efficient manner?}
\end{enumerate}
\vspace{-0.1\baselineskip}

Over the past decade, the \ac{IR} field appears to have reached a consensus that cascading retrieval systems are the answer to this question~\citep{wang2011cascade, liu2017cascade, chen2017efficient, liu2022neural, gao2021rethink, culpepper2016dynamic, clarke2016assessing, asadi2013document, asadi2013effectiveness, zheng2024full, zhang2021learning, qin2022rankflow, guo2022semantic}.
In the cascading paradigm, models are applied in sequence to re-rank the top results of the previous model~\citep{wang2011cascade}.
The order of the sequence follows model scalability, which generally is inversely correlated with ranking effectiveness~\citep{clarke2016assessing}.
In other words, a retrieval model $M_0$ with minimal costs per document is used to create a first ranking (e.g., BM25~\citep{robertson1995okapi} or dense retrieval~\citep{zhan2021optimizing, guo2022semantic}), its top-$K_0$ results are subsequently re-ranked by the next model $M_1$, its top-$K_1$ results are then re-ranked by $M_2$, and so forth. %
The cascading paradigm applies costlier models to fewer documents than cheaper models, this saves costs whilst it also uses the costlier models to improve upon the rankings of cheaper models~\citep{culpepper2016dynamic, pradeep2021expando}.
As a result, a cascading retrieval model can have an efficiency-utility trade-off that none of its models can obtain individually~\citep{wang2011cascade, asadi2013effectiveness, clarke2016assessing, zhang2021learning}.

However, the recent introduction of \acp{LLM} has revealed limitations of the cascade paradigm.
\acp{LLM} are extremely flexible in their application, as they can handle a virtually endless variety of prompts~\citep{team2023gemini, zhao2023survey, chang2024survey, wei2022emergent}.
Previous work found that using \acp{LLM} for predicting relative relevance differences instead of absolute judgements results in substantially better rankings~\citep{qin2024large, liusie2024efficient, yan-etal-2024-consolidating, liu2024aligning}.
For instance, \acfi{PRP}~\citep{qin2024large} has state-of-the-art zero-shot ranking by generating pairwise relevance labels that each indicate whether one document is more relevant than another.
The downside of standard \ac{PRP} is that the number of pairs -and thus the number of \ac{LLM} calls- scales quadratically with the number of documents.
This makes \ac{PRP} extremely costly, even as the final stage in a cascade retrieval system.

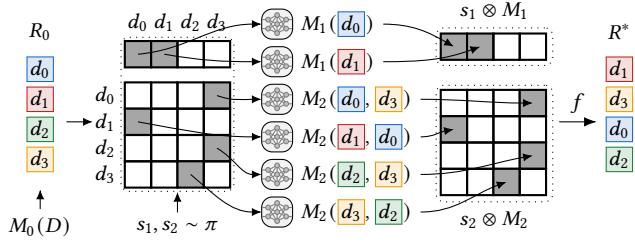
\begin{figure}[t]
    \centering
    \begin{tikzpicture}[
  auto
]

\setlength\fboxsep{1pt}
\tikzstyle{every node}=[font=\small]

\node (m0) at (0.7,-1.4) {$M_0(D)$};

\node (r0) at (0.7, 1.2) [] {$R_0$};
\node (d0) at (0.7, 0.7) [] {\fcolorbox{gBlue}{gBlue!15}{$d_0$}};
\node (d1) at (0.7, 0.3) [] {\fcolorbox{gRed}{gRed!15}{$d_1$}};
\node (d2) at (0.7,-0.1) [] {\fcolorbox{gGreen}{gGreen!15}{$d_2$}};
\node (d3) at (0.7,-0.5) [] {\fcolorbox{gYellow}{gYellow!15}{$d_3$}};

\draw[-{Latex[length=1.3mm,width=1.1mm]}] (m0.north) -- (0.7, -0.9) ;
\draw[-{Latex[length=1.3mm,width=1.1mm]}] (1.05, 0) -- (1.4, 0) ;

\node (s0) at (2,0.9) [thick,draw=black,fill=black!35,minimum height=0.35cm,minimum width=0.35cm] {};
\node (s00) at (2,0.35) [thick,draw=black,minimum height=0.35cm,minimum width=0.35cm] {};
\node (s01) at (2,0) [thick,draw=black,fill=black!35,minimum height=0.35cm,minimum width=0.35cm] {};
\node (s02) at (2,-0.35) [thick,draw=black,minimum height=0.35cm,minimum width=0.35cm] {};
\node (s03) at (2,-0.7) [thick,draw=black,minimum height=0.35cm,minimum width=0.35cm] {};

\node (s1) at (2.35,0.9) [thick,draw=black,fill=black!35,minimum height=0.35cm,minimum width=0.35cm] {};
\node (s10) at (2.35,0.35) [thick,draw=black,minimum height=0.35cm,minimum width=0.35cm] {};
\node (s11) at (2.35,0) [thick,draw=black,minimum height=0.35cm,minimum width=0.35cm] {};
\node (s12) at (2.35,-0.35) [thick,draw=black,minimum height=0.35cm,minimum width=0.35cm] {};
\node (s13) at (2.35,-0.7) [thick,draw=black,minimum height=0.35cm,minimum width=0.35cm] {};

\node (s2) at (2.7,0.9) [thick,draw=black,minimum height=0.35cm,minimum width=0.35cm] {};
\node (s20) at (2.7,0.35) [thick,draw=black,minimum height=0.35cm,minimum width=0.35cm] {};
\node (s21) at (2.7,0) [thick,draw=black,minimum height=0.35cm,minimum width=0.35cm] {};
\node (s22) at (2.7,-0.35) [thick,draw=black,minimum height=0.35cm,minimum width=0.35cm] {};
\node (s23) at (2.7,-0.7) [thick,draw=black,fill=black!35,minimum height=0.35cm,minimum width=0.35cm] {};

\node (s3) at (3.05,0.9) [thick,draw=black,minimum height=0.35cm,minimum width=0.35cm] {};
\node (s30) at (3.05,0.35) [thick,draw=black,fill=black!35,minimum height=0.35cm,minimum width=0.35cm] {};
\node (s31) at (3.05,0) [thick,draw=black,minimum height=0.35cm,minimum width=0.35cm] {};
\node (s32) at (3.05,-0.35) [thick,draw=black,fill=black!35,minimum height=0.35cm,minimum width=0.35cm] {};
\node (s33) at (3.05,-0.7) [thick,draw=black,minimum height=0.35cm,minimum width=0.35cm] {};

\node (sborder) at (2.525, 0.1) [minimum height=2.05cm, minimum width=1.5cm,dotted,draw=black,rounded corners=3pt] {};

\node (s) at (2.525,-1.4) [] {$s_1, s_2 \sim \pi$};
\draw[-{Latex[length=1.3mm,width=1.1mm]}] (s.north) -- (sborder.south) ;

\node (sd0) at (s0.north) [anchor=south] {$d_0$};
\node (sd1) at (s1.north) [anchor=south] {$d_1$};
\node (sd2) at (s2.north) [anchor=south] {$d_2$};
\node (sd3) at (s3.north) [anchor=south] {$d_3$};

\node (sd00) at (s00.west) [anchor=east] {$d_0$};
\node (sd10) at (s01.west) [anchor=east] {$d_1$};
\node (sd20) at (s02.west) [anchor=east] {$d_2$};
\node (sd30) at (s03.west) [anchor=east] {$d_3$};

\pic [local bounding box=m10,scale=0.4] at (3.85,  1.30) {neuralnetwork};
\pic [local bounding box=m11,scale=0.4] at (3.85,  0.80) {neuralnetwork};
\pic [local bounding box=m20,scale=0.4] at (3.85,  0.30) {neuralnetwork};
\pic [local bounding box=m21,scale=0.4] at (3.85, -0.20) {neuralnetwork};
\pic [local bounding box=m22,scale=0.4] at (3.85, -0.70) {neuralnetwork};
\pic [local bounding box=m23,scale=0.4] at (3.85, -1.20) {neuralnetwork};

\draw[-{Latex[length=1.3mm,width=1.1mm]}] (s0.center) to[out=5,in=180] (m10.west) ;
\draw[-{Latex[length=1.3mm,width=1.1mm]}] (s1.center) to[out=-10,in=180] (m11.west) ;
\draw[-{Latex[length=1.3mm,width=1.1mm]}] (s30.center) to[out=-20,in=180] (m20.west) ;
\draw[-{Latex[length=1.3mm,width=1.1mm]}] (s01.center) to[out=-10,in=180] (m21.west) ;
\draw[-{Latex[length=1.3mm,width=1.1mm]}] (s32.center) to[out=-30,in=180] (m22.west) ;
\draw[-{Latex[length=1.3mm,width=1.1mm]}] (s23.center) to[out=-40,in=180] (m23.west) ;

\node (M10) at (m10.east) [anchor=west] {$M_1($\fcolorbox{gBlue}{gBlue!15}{$d_0$}$)$} ;

\node (M11) at (m11.east) [anchor=west] {$M_1($\fcolorbox{gRed}{gRed!15}{$d_1$}$)$} ;

\node (M20) at (m20.east) [anchor=west] {$M_2($\fcolorbox{gBlue}{gBlue!15}{$d_0$}, \fcolorbox{gYellow}{gYellow!15}{$d_3$}$)$} ;
\node (M21) at (m21.east) [anchor=west] {$M_2($\fcolorbox{gRed}{gRed!15}{$d_1$}, \fcolorbox{gBlue}{gBlue!15}{$d_0$}$)$} ;
\node (M22) at (m22.east) [anchor=west] {$M_2($\fcolorbox{gGreen}{gGreen!15}{$d_2$}, \fcolorbox{gYellow}{gYellow!15}{$d_3$}$)$} ;
\node (M23) at (m23.east) [anchor=west] {$M_2($\fcolorbox{gYellow}{gYellow!15}{$d_3$}, \fcolorbox{gGreen}{gGreen!15}{$d_2$}$)$} ;

\node (x0)  at (6.2, 1.00) [thick,draw=black,fill=black!35,minimum height=0.35cm,minimum width=0.35cm] {};
\node (x00) at (6.2, 0.25) [thick,draw=black,minimum height=0.35cm,minimum width=0.35cm] {};
\node (x01) at (6.2,-0.10) [thick,draw=black,fill=black!35,minimum height=0.35cm,minimum width=0.35cm] {};
\node (x02) at (6.2,-0.45) [thick,draw=black,minimum height=0.35cm,minimum width=0.35cm] {};
\node (x03) at (6.2,-0.80) [thick,draw=black,minimum height=0.35cm,minimum width=0.35cm] {};

\node (x1)  at (6.55,  1.00) [thick,draw=black,fill=black!35,minimum height=0.35cm,minimum width=0.35cm] {};
\node (x10) at (6.55, 0.25) [thick,draw=black,minimum height=0.35cm,minimum width=0.35cm] {};
\node (x11) at (6.55,-0.10) [thick,draw=black,minimum height=0.35cm,minimum width=0.35cm] {};
\node (x12) at (6.55,-0.45) [thick,draw=black,minimum height=0.35cm,minimum width=0.35cm] {};
\node (x13) at (6.55,-0.80) [thick,draw=black,minimum height=0.35cm,minimum width=0.35cm] {};

\node (x2)  at (6.90, 1.00) [thick,draw=black,minimum height=0.35cm,minimum width=0.35cm] {};
\node (x20) at (6.90, 0.25) [thick,draw=black,minimum height=0.35cm,minimum width=0.35cm] {};
\node (x21) at (6.90,-0.10) [thick,draw=black,minimum height=0.35cm,minimum width=0.35cm] {};
\node (x22) at (6.90,-0.45) [thick,draw=black,minimum height=0.35cm,minimum width=0.35cm] {};
\node (x23) at (6.90,-0.80) [thick,draw=black,fill=black!35,minimum height=0.35cm,minimum width=0.35cm] {};

\node (x3)  at (7.25,  1.00) [thick,draw=black,minimum height=0.35cm,minimum width=0.35cm] {};
\node (x30) at (7.25, 0.25) [thick,draw=black,fill=black!35,minimum height=0.35cm,minimum width=0.35cm] {};
\node (x31) at (7.25,-0.10) [thick,draw=black,minimum height=0.35cm,minimum width=0.35cm] {};
\node (x32) at (7.25,-0.45) [thick,draw=black,fill=black!35,minimum height=0.35cm,minimum width=0.35cm] {};
\node (x33) at (7.25,-0.80) [thick,draw=black,minimum height=0.35cm,minimum width=0.35cm] {};

\node (s1outborder) at (x1.east) [minimum height=0.5cm, minimum width=1.55cm,dotted,draw=black,rounded corners=3pt,anchor=center] {};
\node (s2outborder) at (x11.south east) [minimum height=1.55cm, minimum width=1.55cm,dotted,draw=black,rounded corners=3pt,anchor=center] {};
\node (s1times) at (s1outborder.north) [anchor=south] {$s_1 \otimes M_1$};
\node (s2times) at (s2outborder.south) [anchor=north] {$s_2 \otimes M_2$};

\draw[-{Latex[length=1.3mm,width=1.1mm]}] (M10.east) to[out=0,in=150]  (x0.center)  ;
\draw[-{Latex[length=1.3mm,width=1.1mm]}] (M11.east) to[out=0,in=200]  (x1.center)  ;
\draw[-{Latex[length=1.3mm,width=1.1mm]}] (M20.east) to[out=0,in=180]  (x30.center)  ;
\draw[-{Latex[length=1.3mm,width=1.1mm]}] (M21.east) to[out=0,in=190]  (x01.center)  ;
\draw[-{Latex[length=1.3mm,width=1.1mm]}] (M22.east) to[out=0,in=190]  (x32.center)  ;
\draw[-{Latex[length=1.3mm,width=1.1mm]}] (M23.east) to[out=0,in=200]  (x23.center)  ;

\node (r0st) at (8.4, 1.2) [] {$R^*$};
\node (d1st) at (8.4, 0.7) [] {\fcolorbox{gRed}{gRed!15}{$d_1$}};
\node (d3st) at (8.4, 0.3) [] {\fcolorbox{gYellow}{gYellow!15}{$d_3$}};
\node (d0st) at (8.4,-0.1) [] {\fcolorbox{gBlue}{gBlue!15}{$d_0$}};
\node (d2st) at (8.4,-0.5) [] {\fcolorbox{gGreen}{gGreen!15}{$d_2$}};

\draw[-{Latex}] (7.6, 0) -- (8.1, 0) node [midway] {$f$};

\end{tikzpicture}
    \vspace{-2\baselineskip}
    \caption{
    Overview of the compound retrieval system described in Section~\ref{sec:method}.
    A first-stage retrieval model $M_0$ retrieves documents to create a first ranking $R_0$; based on their position in $R_0$, the policy $\pi$ decides which documents to apply the pointwise prediction model $M_1$ and pairwise prediction model $M_2$.
    Subsequently, the predictions of $M_1$ and $M_2$ are only gathered where activated and combined into a final ranking $R^*$ using the score aggregation function $f$. 
    }
    \label{fig:overview}
    \vspace{-1.5\baselineskip}
\end{figure}

Inspired by this shortcoming and because the cascading paradigm has rarely been questioned in the \ac{IR} field, we reconsider the question stated in the first paragraph. 
Our first contribution is introducing the concept of a \emph{compound retrieval system} that encompasses any retrieval system that applies multiple predictions models.
By keeping our definition broad and avoiding assumptions, we aim to explore the largest design space possible.
We argue that compound retrieval system design depends on two essential questions:
\vspace{-0.1\baselineskip}
\begin{enumerate}[leftmargin=*,label=(\roman*)]
    \item \emph{What predictions should the component models make?}
    \item \emph{How should predictions be combined to construct a ranking?}
\end{enumerate}
We note that the cascade paradigm has very specific answers: model predictions should be made on documents in the top-$K$ of previous models; and predicted scores should be used to re-rank the same top-$K$~\citep{wang2011cascade}.
Therefore, cascading retrieval systems are a subset of compound retrieval systems, which is thus a generalization.

We argue that the answers to these questions should depend on both the costliness and usefulness of the component model predictions, in addition to the efficiency constraints and ranking quality needs one has for their retrieval system.
Accordingly, instead of providing prescriptive answers to these questions, our second contribution is a framework that optimizes the design of a compound retrieval system \emph{automatically}.
We break the system design down in a policy that decides what model predictions to gather, and a prediction aggregation function to construct a ranking out of the gathered predictions.
Given a differentiable loss function, our framework can then optimize the system.
For example, given a linear interpolation of a ranking utility and prediction costs, the framework will optimize the design for the corresponding effectiveness-efficiency trade-off.
Importantly, it is capable of finding novel system designs that fall outside the cascading paradigm, e.g., by combining multiple predictions in a single re-ranking step, or by treating documents differently depending on their rank in the previous ranking.

In our experiments, we optimize a compound retrieval system that combines classic BM25 retrieval with pointwise and pairwise \ac{LLM} predictions.
We perform optimization both in a supervised setting where relevance labels are available, and in a self-supervised setting where the goal is to reconstruct the most-costly \ac{PRP} ranking.
Our efficiency target is to limit the number of \ac{LLM} calls, we vary its importance to construct effectiveness-efficiency trade-offs of our model.
Our results reveal that our optimized compound retrieval systems have better trade-off curves than cascading systems.
In particular, we can closely match the nDCG of a top-1000 \ac{PRP} re-ranker with only a fraction of its \ac{LLM} calls, regardless of whether relevance labels are available for supervision.
An analysis of our optimized designs reveal a large variety of novel strategies depending on the desired effectiveness-efficiency trade-off.
For example, on the extremes, they default to BM25 to avoid costs or to PRP to maximize effectiveness; for trade-offs in the middle, they gather pointwise and pairwise predictions in targeted patterns based on the BM25 ranking.
This bolsters our claim that compound retrieval system designs should adapt to its components and the desired system properties, without being constrained to the cascade paradigm.

To the best of our knowledge, this is the first work to consider alternatives to the existing cascade paradigm.
Our work reveals a high potential for designs outside this paradigm, and for our optimization framework for finding such designs.
We hope that it provides a starting point for a novel research direction that further explores compound retrieval system design and optimization.

\section{Definition: Compound Retrieval Systems}
\label{sec:compounddefinition}

Inspired by compound A.I.\ systems~\citep{zaharia2024shift}, we define compound retrieval systems as any retrieval system that constructs a ranking by using multiple prediction models.
Conversely, a non-compound retrieval system uses a singular monolithic ranking model~\citep{zaharia2024shift, wang2011cascade}.
Components are prediction models that give scores to individual or sets of documents, these can be machine learned models but do not need to be.
We denote the set of components of a system as $\{M_0,M_1,\ldots\}$ and $\phi$ as the aggregation function that defines how the components interact with each other to construct rankings.
Thus, we use $\phi(M_0,M_1,\ldots)$ to denote a compound retrieval system.

\section{Background}
\label{sec:background}
\subsection{Cascade ranking systems}
There is a seeming consensus on the cascading paradigm for the trade-off between effectiveness and efficiency in the IR field.
It underlies virtually every advancement in retrieval and ranking of the past decade, as (almost) every model is ultimately applied inside a cascade retrieval system~\citep{wang2011cascade, liu2017cascade, chen2017efficient, liu2022neural, gao2021rethink, culpepper2016dynamic, clarke2016assessing, asadi2013document, asadi2013effectiveness, zheng2024full, zhang2021learning, qin2022rankflow, guo2022semantic}.

Cascading ranking systems balance the quality of their rankings with their computation costs by applying several ranking models in sequence~\citep{wang2011cascade}.
Generally, a first-stage retrieval model like BM25~\citep{robertson1995okapi} creates the first ranking of the collection, of which only a top-$K_0$ is used.
Let $M_0$ be a first-stage ranking model and its top-$K_0$ ranking:
\begin{equation}
\begin{split}
&R_0 = [d_1, d_2, d_3, \ldots d_{K_0}]
\\
&\text{s.t. } \forall 1 \leq i \leq K_0, \forall d \in D \setminus [d_1,\ldots,d_{i-1}],
\quad M_0(d_i) \geq M_0(d).
\end{split}
\end{equation}
It is important that $M_0(d)$ is scalable so that $R_0$ can be computed over a document collection $D$ with reasonable costs.
There is extensive literature on efficient first-stage retrieval~\citep{guo2022semantic, tonellotto2018efficient, moffat1996self, petri2013exploring}.

Subsequently, the next ranking model $M_1$ is applied to re-rank $R_0$ and produce a shorter ranking $R_1$.
Because $R_0$ does not include the entire collection ($|D| \gg K_0$), the costs of $M_1$ can be higher as it is applied to much fewer documents.
This process repeats for several stages that each apply a more costly model to fewer documents~\citep{culpepper2016dynamic}.
In our notation, the ranking at stage $i$ is:
\begin{equation}
\begin{split}
R_i = [d_1, d_2, \ldots,& {} d_{K_i}]
 \, \text{s.t. }    \forall 1\! \leq\! j \!\leq \!K_i, \, \forall d \in\! D \!\setminus\! [d_1,\ldots,d_{j-1}],
\\ &\quad\,
d_j \in R_{i-1} \land (d \not\in R_{i-1} \lor M_i(d_j) \geq M_i(d)).
\end{split}
\end{equation}
In the framing of our work, the cascading ranking model is a subclass of the compound retrieval system, where $\phi_\text{casc}(M_0,\ldots,M_N) = R_N$, i.e., the aggregation function $\phi_\text{casc}$ only lets components interact through sequential top-$K$ re-ranking operations~\citep{wang2011cascade}.

\subsection{Pairwise relevance prompting}
\acp{LLM} are better at prediction relevance differences than absolute relevance values~\citep{qin2024large, liusie2024efficient, liu2024aligning}.
Thus, using an \ac{LLM} to predict whether one document is more relevant than another leads to better rankings than using predictions about individual documents~\citep{qin2024large, yan-etal-2024-consolidating}.
Accordingly, the \acfi{PRP} approach~\citep{qin2024large} uses an LLM with a pairwise prompt to predict the probability that one document should be ranked before another:
\begin{equation}
    M_\text{PRP}(d_1, d_2) = \hat{P}_\text{LLM}(\upsilon_{d_1} > \upsilon_{d_2}).
\end{equation}
To turn pairwise predictions into a valid ranking, \ac{PRP} ranks documents by their predicted win-rate.
Thus, formally, \ac{PRP} uses the following scoring function to re-rank the first-stage ranking $R_0$:
\begin{equation}
    M_\text{PRP}(d \mid R_0 \setminus d)
    =
    \frac{1}{2}\sum_{d'\! \in R_0 \setminus d} M_\text{PRP}(d'\!, d)
    + 1 - M_\text{PRP}(d, d').
    \label{eq:prpaggregation}
\end{equation}
Due to the costly nature of LLM predictions and since the number of pairs grows quadratically with the number of documents ($|R_0|^2 - |R_0|$), previous work applies \ac{PRP} only to re-rank a top-$K$ from a first stage ranker ($M_0$)~\citep{qin2024large, yan-etal-2024-consolidating}.

\citet{qin2024large} propose strategies for reducing the number of pairs to consider, e.g., by mimicking sorting algorithms.
However, this creates sequential processes that cannot be parallelized, for their impracticality, this work will not consider such approaches.

\section{Method: A Framework for Optimizing \phantom{Method:} Compound Retrieval Systems}
\label{sec:method}

This section proposes a compound retrieval system (Section~\ref{sec:compoundsystem}) that is a generalization of a cascading system~(Section~\ref{sec:generalization}).
Subsequently, we introduce our optimization framework that learns how predictions can be aggregated (Section~\ref{sec:method:ranking}), and which predictions to gather (Section~\ref{sec:method:costs}), we also show how our system can be made deterministic (Section~\ref{sec:method:deterministic}).
Finally, we summarize our approach (Section~\ref{sec:method:summary}). A visual overview of our system is shown in Figure~\ref{fig:overview}.

\subsection{A compound retrieval system}
\label{sec:compoundsystem}

In this method section, we describe a compound retrieval system with three components $\{M_0, M_1, M_2\}$: $M_0$ is a first-stage retrieval model, $M_1$ is a pointwise prediction model and $M_2$ is a pairwise prediction model.
Our framework can easily be extended for other choices of component models (see Section~\ref{sec:exp:components} and~\ref{sec:method:extensions}).

Formally, our compound retrieval system describes a new ranking function that is based on the predictions of its underlying components.
In contrast with cascading systems, our compound system can choose what $M_1(d) \in \mathds{R}$ and $M_2(d, d') \in \mathds{R}$ predictions it wishes to gather.
Let $\pi$ be the selection policy of our compound system, i.e, a probability distribution over the possible selections of the predictions that can be gathered.
The vector $s_1$ indicates the selections made for $M_1$ and the matrix $s_2$ the selections for $M_2$:
\begin{equation}
    s_1, s_2 \sim \pi, \qquad s_1 \in \{0,1\}^{K_0}, \qquad s_2 \in \{0,1\}^{K_0 \times K_0}.
    \label{eq:selectionpolicy}
\end{equation}
Thus, $s_1$ is a binary vector of size $K_0$ whose indices correspond to the ranking of the first-stage model $M_0$, e.g., $s_1[1] = 1$ indicates that the prediction of $M_1$ of the document ranked first by $M_0$ is selected. %
Similarly, $s_2$ is a binary matrix of size $K_0 \times K_0$ representing every ordered pair of documents corresponding to the first-stage ranking $R_0$,\footnote{Pairs in $s_2$ are ordered since for \ac{PRP} the order matters: $M_2(d,d') \not= M_2(d'\!,d)$.} %
e.g., $s_2[4,8] = 1$ indicates that the $M_2$ prediction for the $4$th and $8$th documents, as ranked by $M_0$ are selected.
To keep notation brief, we indicate whether a document or a pair is selected by:
$s_1(d) = s_1[R_0^{-1}(d)]$ and $s_2(d,d') = s_2[R_0^{-1}(d),R_0^{-1}(d')]$,
where $R_0^{-1}(d)$ indicates the rank of $d$ in ranking $R_0$.%
\footnote{Hence $R_0^{-1}(d)$ is the inverse of $R_0[i]$ which gives the document at rank $i$ in $R_0$.}
Thus, $s_1(d)$ indicates whether $M_1(d)$ is selected and $s_2(d,d')$ whether $M_2(d,d')$ is.

Predictions that are not selected should be distinguishable from zero predictions, i.e., $s_1(d) = 1$ and $M_1(d) = 0$ should be differentiated from $s_1(d) = 0$ and $M_1(d) \not= 0$.
For this reason, we introduce the masking operator $\otimes$ which outputs tuples that contain both selection indicators and masked predictions:
\begin{equation}
\begin{split}
    (s_1 \otimes M_1)(d) &= (s_1(d), \, s_1(d) \cdot M_1(d)), \\
    (s_2\otimes M_2)(d, D) &= ((s_2(d,d'), \, s_2(d,d')\cdot M_2(d, d')) \! : d'\! \in D).
\end{split}
\label{eq:maskedinput}
\end{equation}

Finally, in order to construct a ranking, we use a function $f$ that takes three inputs: $(R_0^{-1}(d), (s_1 \otimes M_1)(d), (s_2 \otimes M_2)(d, R_0))$.
This function produces the scores to determine the final ranking, and thus, it can be interpreted as a final ranking function:
\begin{equation}
    M^*(d \mid f, s) = f(R_0^{-1}(d), (s_1 \otimes M_1)(d), (s_2 \otimes M_2)(d, R_0)),
\label{eq:finalrankingfunction}
\end{equation}
where the final ranking is the corresponding re-ranking of $R_0$:
\begin{equation}
\begin{split}
R^*(\pi, {} & {} f, s) = [d_1, d_2, d_3, \ldots d_{K_i}]
\\
& \text{s.t. } \forall 1 \leq j \leq K_i, \; \forall d \in D \setminus [d_1,\ldots,d_{j-1}],
\\ &\quad\;\,
d_j \in R_0 \land (d \not\in R_0 \lor M^*(d_j \mid f, s) \geq M^*(d \mid f, s)).
\end{split}
\end{equation}
Accordingly, the component aggregation function of our compound retrieval system is defined as:
\begin{equation}
    \phi_\text{comp}(M_0, M_1, M_2 \mid \pi, f) = R^*(\pi,f),
\end{equation}
and determined by selection policy $\pi$ and score aggregator $f$.

Correspondingly, the choice of $\pi$ and $f$ is highly important as it completely defines the system behavior.
For this work, we propose a method that searches for the $\pi$ and $f$ by optimizing a given trade-off between efficiency and effectiveness.
Formally, we aim to minimize a linear interpolation of a ranking loss and the cost of the system, for given $\alpha \in [0,1]$ the loss for our compound system is:
\begin{equation}
    \mathcal{L}_\text{comp}(f, \pi) = \alpha \mathcal{L}_\text{ranking}(\pi,f) + (1-\alpha)\mathcal{L}_\text{cost}(\pi).
    \label{eq:loss}
\end{equation}
Our cost is the expected number of LLM calls:
\begin{equation}
\mathcal{L}_\text{cost}(\pi) = \mathds{E}_{(s_1,s_2)\sim\pi}\bigg[
\sum_{d \in R_0} s_1(d) + \sum_{d \in R_0}\sum_{d' \in R_0} s_2(d,d')
\bigg].
\end{equation}
Section~\ref{sec:method:costs} further details how we optimize $\pi$.
For the ranking loss $\mathcal{L}_\text{ranking}(\pi,f)$, we utilize existing \ac{LTR} loss functions for supervised learning (from labels)~\citep{jagerman2022rax} and self-supervised learning (ranking distillation)~\citep{qin2023rd}, Section~\ref{sec:method:ranking} discusses how these can be applied to our compound retrieval systems framework.

\subsection{Comparison with cascades and PRP}
\label{sec:generalization}
As laid out in Sections~\ref{sec:compounddefinition} and~\ref{sec:background}, cascading retrieval systems are particular instances of compound retrieval models.
In our framework, we can choose $\pi$ and $f$ such that the system becomes equivalent to a specific cascading system.
For example, for a cascade re-ranking with $M_1$, we choose $\pi$ as a deterministic policy that selects all pointwise predictions, and a ranking function $f$ that outputs them:\footnote{We are describing equivalence with a cascade model with a single re-ranking step, but our framework is easily extended to multiple re-ranking steps, see Section~\ref{sec:method:extensions}.}
\begin{equation}
    s_1 = \mathbf{1}, \quad s_2 = \mathbf{0}, \quad f(\cdots) = M_1(d).
    \qquad\quad \text{(Cascade)}
\end{equation}
Similarly, for PRP, we choose $\pi$ to only select all pairwise predictions and rank with the PRP aggregation function (Eq.~\ref{eq:prpaggregation}):
\begin{equation}
    s_1 = \mathbf{0}, \quad s_2 = \mathbf{1}, \quad f(\cdots) = M_2(d, R_0).
    \qquad\qquad \text{(PRP)}
\end{equation}
Moreover, our framework also allows us to ignore all predictions from $M_1$ and $M_2$ and leave the first-stage ranking unchanged: %
\begin{equation}
    s_1 = \mathbf{0}, \quad s_2 = \mathbf{0}, \quad f(\cdots) = - R_0^{-1}(d) \equiv M_0(d).
    \; \text{(First-Stage)}
\end{equation}
Furthermore, our framework can also capture changes in parameters for these systems, e.g., using to re-rank a smaller top-$K$ by only selecting predictions of that top-$K$ and scoring accordingly.

This means our compound framework provides a generalization of all of these three approaches.
Consequently, our optimization can choose from any of these approaches, e.g., PRP when high costs are allowed ($\alpha=1$), first-stage retrieval when no costs are allowed ($\alpha=0$) or a pointwise cascade re-ranking for a trade-off in between.
Importantly, there are also many other possible interactions between components that fit in our framework.
Thus, our framework can produce many compound retrieval systems that do no match existing cascading retrieval systems.
Since there is no precedent, it is unclear what these interactions should be like, and our optimization framework might discover them for the first time.

To the best of our knowledge, this is the first work that explores these alternative compound retrieval systems.

\subsection{Learning how to aggregate predictions}
\label{sec:method:ranking}
There are many choices possible for the scoring aggregation function $f$; for this work, we propose an extremely simple model as it allows us to interpret its behavior and ensures that it can be implemented in a large scale practical setting.

We break down the scoring function $f$ into three parts: $f = (f_0, f_1, f_2)$, corresponding to the three component models.
Accordingly, each deals with the rank in the first-stage ranking and the (masked) outputs of one of the components (Eq.~\ref{eq:maskedinput} and~\ref{eq:finalrankingfunction}):
\begin{equation}
f_0 : \mathds{Z} \rightarrow \mathds{R}, \quad
f_1 : \mathds{Z}_2 \times \mathds{R} \times \mathds{Z}  \rightarrow \mathds{R}, \quad
f_2 : \mathds{Z}_2 \times \mathds{R} \times \mathds{Z}^2  \rightarrow \mathds{R}.
\nonumber
\end{equation}

To start, $f_0$ takes as input the rank of document according to the first-stage ranker $M_0$ as $r$ and returns the variable $A_r$:
\begin{equation}
    f_0(r) = A_{r}.
    \label{eq:f0}
\end{equation}
This is a \emph{default} score: the score a document gets when no related predictions of $M_1$ or $M_2$ are used.
We note that this score only depends on the rank of document in the first-stage ranking $R_0$.

The second function $f_1$ has as input a selection indicator $s \in \{0,1\}$, $m$ the (masked) prediction of a document given by $M_1$ (i.e., $M_1(d)$) and the rank in the first stage ranking $r$, its output is based on two variables $B_r$ and $C_r$ which are unique per $r$:
\begin{equation}
    f_1(s, m, r)
    = s\cdot(B_{r} + C_{r}\cdot m).
    \label{eq:f1}
\end{equation}
In other words, $f_1$ provides a zero score if no selection is made ($s=0$), if a selection is made ($s=1$) the score starts with value $B_r$ and adds $C_r$ proportional to the prediction $m$ ($M_1(d)$).
If $m\in[0,1]$, this is a linear interpolation between $B_r$ and $B_r+C_r$.

The final function $f_2$ also takes the selection indicator $s \in \{0,1\}$ as input, along with $m$, the masked prediction for a document pair given by $M_2$ (i.e., $M_2(d_1,d_2)$) and the ranks of both document in the first stage ranking $r_1$ and $r_2$.
Its scoring is also based on two variables $B_{r_1,r_2}$ and $C_{r_1,r_2}$ which are unique per rank pair $(r_1,r_2)$:
\begin{equation}
    f_2(s, m, r_1, r_2)
    = s\cdot(B_{r_1,r_2} + C_{r_1,r_2}\cdot m).
    \label{eq:f2}
\end{equation}
Similarly, it is zero if the prediction is not selected; when selected, it starts with $B_r$ and $C_r$ is added in proportion to the prediction.

The ranking score of our final ranking model is the addition of all predictions related to a document transformed by our functions, this includes a summation over all pairs where $d$ is the first document:
\begin{align}
    M^*&(d \mid f, s) 
    = 
    f(R_0^{-1}(d), (s_1 \otimes M_1)(d), (s_2 \otimes M_2)(d, R_0))
    \nonumber\\  = {} & 
    f_0(R_0^{-1}(d))
    + f_1(s_1(d), s_1(d) \cdot M_1(d), R_0^{-1}(d))
    \\
    & + \!\!\! \sum_{d' \in R_0 \setminus d} \!\!\!\! f_2(s_2(d,d'), s_2(d,d') \cdot M_2(d,d'),R_0^{-1}(d),R_0^{-1}(d')).
    \nonumber
\end{align}
Our ranking model is completely differentiable w.r.t.\ the variables $A$, $B$ and $C$, and therefore, $f$ can be optimized with standard \ac{LTR} methods~\citep{oosterhuis2021computationally, jagerman2022rax}.
However, since $s$ are sampled binary variables, it does not work with standard stochastic gradient descent; the next section describes how we optimize its underlying selection policy.

\subsection{Learning where to apply components}
\label{sec:method:costs}

The optimization of the selection policy $\pi$ is more complicated as the sampling operation is not differentiable.
Many solutions to this issue exist, most notably one can compute its gradient through estimating it based on many samples, i.e., with REINFORCE~\citep{williams1992simple}.
However, this often has high variance which can prevent optimal convergence, and it is computationally very costly.
For this reason, we apply the cheaper and more stable straight-through-estimator~\citep{bengio2013estimating}; thereby, we estimate the gradient of each selection to be its probability:
\begin{equation*}
\begin{split}
\frac{\delta s_1(d)}{\delta \pi(s_1(d)\! = \!1)} \!=\! \pi(s_1(d)\! =\! 1),\;
\frac{\delta s_2(d,d')}{\delta \pi(s_2(d,d')\! =\! 1)} \! =\! \pi(s_2(d,d')\! =\! 1).
\end{split}
\end{equation*}
This is a biased estimate of the actual gradient, nevertheless, for most purposes its is accurate enough whilst also providing simplicity and stability at low computational costs.

As a result, we can now apply stochastic gradient descent to our entire compound retrieval system from the loss in Eq.~\ref{eq:loss}.
For the policy $\pi$, what happens is that if the \ac{LTR} loss indicates that the ranking score of a document should increase, the accompanying gradient will increase the probabilities of all selections that would result in an increase in score, or decrease probabilities if that results in a score increase.
Vice versa, if  document score should decrease, probabilities are pushed in the other direction.
Similarly, the cost part of the loss will produce a gradient that decreases the probability of each selection, according to how much a selection would increase costs.
Together, the interpolation of the ranking and costs losses results in a gradient that decreases most selection probabilities except for those that lead to an increase in ranking performance that outweighs the added selection cost.

\subsection{Finding a deterministic policy}
\label{sec:method:deterministic}
We have described our compound retrieval system and how its score aggregation function $f$ and selection policy $\pi$ can be optimized.
So far, we have let $\pi$ be a probabilistic policy, as this makes its optimization much easier.
However, in practice, it is often more beneficial to have a deterministic policy, as these generally allow for substantially more efficient implementations.
For this reason, we add an additional step at the end of our optimization where we take a number of selection samples from $\pi$ and select the sample with the lowest loss on a validation set.
Let $s_1^*,s_2^*$ indicate our best sample, the selection policy is then changed to a deterministic policy that always selects this sample: $\pi(s_1=s_1^*,s_2=s_2^*) = 1$.

\subsection{Efficient representation and computation}
\label{sec:matrixnotation}
In our description of our score aggregation function in Section~\ref{sec:method:ranking}, we noted that it was chosen for its simplicity that enables it to be implemented for large scale settings.
We believe this is the case because it can be computed solely via basic matrix operations:

To start, all the model variables of our system in Eq.~\ref{eq:f0}, Eq.~\ref{eq:f1} and Eq.~\ref{eq:f2} can be captured in one vector and two matrices where the top rows of the matrices are for $f_1$ and the rest for $f_2$.
Similarly, a sample of selections can be represented in a single matrix, with the top row indicating the selection of pointwise predictions:
\begin{equation}
    A \in \mathds{R}^{K_0}, \;\; B, C \in \mathds{R}^{K_0 \times (K_0+1)},
    \quad
    S =
    \begingroup
    \setlength\arraycolsep{3pt}
    \begin{bmatrix}
        s_1\\
        s_2
    \end{bmatrix}
    \endgroup
    \in \mathds{Z}_2^{(K_0+1) \times K_0}.
    \label{eq:selectionmatrix}
\end{equation}
Accordingly, we can gather all model predictions in a single matrix $X \in \mathds{R}^{(K_0+1) \times K_0}$ following the same placement pattern:
\begin{equation*}
    X \!=\!
    \begingroup
    \setlength\arraycolsep{0.75pt}
    \begin{bmatrix}
        M_1(R_0[1]) & M_1(R_0[2]) & \cdots & M_1(R_0[K_0])\\
        M_2(R_0[1],R_0[1]) & M_2(R_0[1],R_0[2]) & \cdots & M_2(R_0[1],R_0[K_0])\\
        \cdots & \cdots & \cdots & \cdots \\
        M_2(R_0[K_0],R_0[1]) & \cdots & \cdots & M_2(R_0[K_0],R_0[K_0])
    \end{bmatrix}
    \endgroup
    .
\end{equation*}
During inference, $X$ should only be populated with predictions that were selected in $S$.
However, during training, $X$ can be pre-computed once and stored, subsequently, we do not have to call the underlying component models during optimization.

Finally, with the matrix representations and the element-wise multiplication $\odot$, our ranking function can be re-formulated to:
\begin{equation}
    M^*( D \mid f) 
    =
    A + B  S + C(S \odot X).
    \label{eq:finalrankingscorematrix}
\end{equation}

Due to the high number of variables, we recommend using an underlying model to produce them.
In our experiments, we use a simple neural network and give it as input matrix indices to produce our variables $A, B, C$ and our selection probabilities $\pi$.
Accordingly, back-propagation can be used to optimize the neural network.
One can also learn the variables directly, but we found that using an underlying model gives more regularity in model behavior which leads to better generalization and little overfitting.
Because $A$, $B$, $C$ and $\pi$ can be pre-computed, there is no added costs during inference.

\subsection{Summary}
\label{sec:method:summary}

This section has described a framework for optimizing a compound retrieval system that can utilize the predictions of three component models: the first-stage retrieval model $M_0$, the pointwise relevance prediction model $M_1$, and the pairwise relevance prediction model $M_2$.
For clarity, we summarize the system that results from our optimization by describing each step that it performs for inference:
\begin{enumerate}[leftmargin=*]
    \item Compute a first ranking $R_0$ according to component $M_0$.
    \item Sample a selection indicator matrix $S$  from the selection policy $\pi$ (Eq.~\ref{eq:selectionmatrix}); selection probabilities can be pre-computed and kept ready in memory.
    We note that selections are based on document positions in the first ranking $R_0$, i.e., selections have the form of \emph{the relevance prediction for the $i$th document in $R_0$}, or \emph{the pairwise prediction for $i$th and $j$th documents in $R_0$}.
    \item Obtain the relevance predictions \emph{only} for the selected single documents $M_1(d)$ and pairs $M_2(d,d')$.
    Fill the $X$ matrix with the selected predictions (the rest remains empty).
    \item Compute a ranking score for each document through matrix operations on $S$ and $X$ according to Eq.~\ref{eq:finalrankingscorematrix}.
    All other variables for this operation can be pre-computed and kept in memory.
    \item Return the final ranking by sorting according to the ranking scores (descendingly).
\end{enumerate}
The main difference for optimization is that for gradient computation all predictions are needed, thus $X$ has to be filled completely.
Fortunately, $X$ can be pre-computed for a static dataset.
Gradient descent is applied to a combination of a ranking loss on the final ranking and a cost loss on the selection policy (Eq.~\ref{eq:loss}).

As discussed in Section~\ref{sec:generalization}, an important property of our compound retrieval system is that it is a generalization of its component models.
This means that its parameters can be set to give identical rankings to any of its components, or any top-$K$ re-ranking of its components (if $K$ < $K_0$).
In other words, it can perform any top-$K$ re-ranking with the pointwise model $M_1$, or with the pairwise model $M_2$, as a cascade retrieval model would do.
It can also simply provide the first-stage retrieval ranking of $M_0$.
Thus, if the optimization is successful, our compound retrieval system should provide an effectiveness-efficiency trade-off that is at least as good as any of its components.
Since it can also capture behavior that none of the components can, it has a high potential of further improvements.

\section{Method: Novel Ranking Losses}
\label{sec:methodrankinglosses}

Our proposed framework aims to maximize ranking utility under efficiency constraints (see the trade-off loss in Eq.~\ref{eq:loss}). Let $\omega_i$ be a weight per rank $i$ and $\upsilon_d$ a relevance label  per document $d$,
our utility function is a sum over the relevance label multiplied with the weight at each rank, e.g., for \ac{DCG}~\citep{jarvelin2002cumulated}:
\begin{equation}
    U(y, \upsilon) = \sum_{i=1}^{|y|} \omega_i \upsilon_{y_i},
    \qquad
    \omega_i^\text{DCG@K}  = \frac{\mathds{1}[i \leq K]}{\log_2(i + 1)}
    .
    \label{eq:DCG}
\end{equation}

\subsection{Ranking distillation by lower bounding}
\label{sec:rankingdistillation}

Ranking distillation losses penalize the difference between two rankings to optimize a ranking model to behave similarly to another~\citep{tang2018ranking}.
We wish to apply such a ranking loss in our trade-off loss in Eq.~\ref{eq:loss} for settings where relevance labels are unavailable.
However, to optimize the trade-off there needs to be a correspondence between the  distillation loss and ranking utility, existing distillation losses are not based on metrics, and thus, miss this property~\citep{qin2023rd}.

We propose a novel ranking distillation approach that is a bound on the difference in utility between two rankings~\citep{gupta2024practical}.
Let $y'$ be a new ranking and $y$ be the original ranking, we aim to bound the maximum loss in utility when switching from $y$ to $y'$:
\begin{equation}
    \mathcal{L}(y, y') \leq  - \max_{\upsilon} U(y, \upsilon) - U(y'\!\!, \upsilon) =  - \min_{\upsilon} U(y'\!\!, \upsilon) - U(y, \upsilon).
\end{equation}
For ease of notation, we define $\omega_{d,y}$ as the weight at the rank of document $d$ in ranking $y$, with $\omega_{d,y} = 0$ if $d$ is not in $y$.
Furthermore, we assume that relevance is always between zero and some maximum value $V$: $\upsilon_{d} \in [0,V]$.
This allows us to make the following derivation that leads to our novel loss $\mathcal{L}(y, y')$:
\begin{align}
    &\min_{\upsilon} U(y'\!\!, y) - U(y, y)
    = \min_{\upsilon} \sum_{d \in y \cup y'} (\omega_{d,y'} - \omega_{d,y}) \upsilon_{d} 
     \label{eq:step:zerorel} \\[-1ex]
    &\quad= \min_{\upsilon} \sum_{d \in y} \min(0, \omega_{d,y'} - \omega_{d,y}) \upsilon_{d}
    = V \!\!\sum_{d \in y} \min(0, \omega_{d,y'} - \omega_{d,y})
    \nonumber \\[-1ex]
    &\quad\propto  \sum_{d \in y} \min(0, \omega_{d,y'} - \omega_{d,y})
     \coloneq  -\mathcal{L}(y, y'). 
    \nonumber
\end{align}
We note that the step after (\ref{eq:step:zerorel}) is valid because $\upsilon_d = 0$ is possible, hence any positive difference of $\omega_{d,y'} - \omega_{d,y}$ is always cancelled out by a zero relevance label in the minimization.
For similar reasons, we can discard all documents that are in $y'$ but not in $y$.

We see that our loss penalizes all documents that are ranked lower in $y'$ than in the original $y$, but gives no reward for ranking documents higher.
Penalties are based on differences in metric weights.
Thus, the loss is minimized when the rankings are identical ($\mathcal{L}(y, y) = 0$).
Importantly, $\mathcal{L}(y, y')$ stays within the range of possible DCG differences, and is proportional to the largest possible loss in DCG between the rankings.
In theory, this makes it a good choice for optimizing our effectiveness-efficiency trade-off.

\subsection{Optimizing ranking metrics with cutoffs}
\label{sec:methodrankinglosses:cutoffs}
For our experiments, we adopt the approach of \citet{qin2010general} by applying differentiable approximation of the ranks of documents: $\widetilde{\text{rank}}(d,y) \approx \text{rank}(d,y)$.
These are straightforwardly used to approximate the metric weights of documents, e.g., for DCG:
\begin{equation}
    \tilde{\omega}_{d,y}^\text{DCG} = \log_2(\widetilde{\text{rank}}(d,y) + 1)^{-1}
    \approx \omega_{d,y}^\text{DCG}.
\end{equation}
However, for ranking metrics with cutoffs, e.g., DCG@K, an additional approximation is needed for the cutoff ($\mathds{1}[\text{rank}(d,y) \leq K]$).
\citet{jagerman2022rax} apply a sigmoid function to approximate cutoffs $\big(\mathds{1}[\text{rank}(d,y) \leq K] \approx \text{sigmoid}(\widetilde{\text{rank}}(d,y) - k)\big)$, but we find that its gradient diminishes too quickly.
As a result, if a document's (approximated) rank is far from the cutoff, the gradient is virtually zero, and thereby, gradient descent is ineffective.

As an alternative, we propose the following approximation which has a gradient that diminishes much slower:
\begin{equation}
    \mathds{1}[i \leq K]
    \approx
    \max(i - K + 1, 1)^{-1}.
\end{equation}
Additionally, since the progression of the weight function beyond the cutoff rank has no effect on the ranking metric, we prevent it from affecting our approximation.
This is done by making $K$ the maximum of the input to the discounting function, together with our cutoff approximation our differentiable document weight is:
\begin{equation}
    \omega_{d,y}^\text{DCG@K} \approx \tilde{\omega}_{d,y}^\text{DCG@K}
    = \frac{
    \max(\widetilde{\text{rank}}(d,y) - K + 1, 1)^{-1}
    }{\log_2(\min(\widetilde{\text{rank}}(d,y), K) + 1)}.
\end{equation}

\section{Experimental Setup}

Our experiments compare the effectiveness-efficiency trade-offs of compound and cascade systems. %
Our implementation is available at: \url{https://github.com/google-deepmind/compound_retrieval}.

\subsection{Dataset}

For our experiments we use the TREC-DL (2019-2022) dataset~\citep{craswell2021trec}.
It consists of documents and queries together with human-annotated relevance judgments for query-document pairs.
We randomly split the queries to create validation and test sets of 50 queries each, the remaining queries are the training set.
For cross-validation, 50 random splits are generated to repeat our experimental runs over.
Our setting is a top-1000 re-ranking task with BM25 as the first stage-ranker, thus, per query, only documents in the top-1000 of BM25 are used, the rest are discarded.

\subsection{LLM prompts and predictions}

Our compound retrieval model is built on the first-stage ranker BM25~\citep{robertson1995okapi} and the \emph{Gemini 1.5 Flash} \ac{LLM}~\citep{team2023gemini}.
We utilize both a pointwise and a pairwise prompt and use \emph{scoring mode} to obtain the probabilities of predictions.
For the pointwise prompt, we use the binary relevance prompt, first introduced by~\citet{liang2022holistic}, using the same prompt as the one described in Figure 1a of~\citep{qin2024large}:
\begin{equation}
\text{LLM}_\text{point}(d) = \hat{P}_{\text{LLM}}(\text{`\emph{Yes}'} \mid \text{prompt}, d).
\end{equation}
We normalize the predicted probability based on the model logits for the answers `\emph{Yes}' and `\emph{No}', such that the probabilities of these two answers sum to one.
Similarly, we use the pairwise PRP prompt described in Figure 2 of~\cite{qin2024large}.
\begin{align}
\text{LLM}_\text{pair}(d, d') &= \hat{P}_{\text{LLM}}(\text{`\emph{Passage A}'} \mid \text{prompt}, d, d').
\end{align}
Importantly, passage A is always presented first (corresponding to $d$) and passage B second ($d'$), and again, we normalize the probabilities based on the logits for the phrases `\emph{Passage A}' and `\emph{Passage B}'.

\subsection{Component models}
\label{sec:exp:components}
For increased performance, we provide our system with different transformations of each prediction, this makes it an extension of the system described in Section~\ref{sec:method}.
In the terms of our formal framework, this results in eight component models in our system:
\begin{align}
    M_0(d) &= \text{BM25}(d),  \\
    M_1(d) &= \text{LLM}_\text{point}(d),
    \hspace{1.45cm}
    M_2(d) = \text{round}(M_1(d)), \nonumber \\
    M_3(d,d') &= \text{LLM}_\text{pair}(d, d'),
    \hspace{0.8cm}
    M_4(d,d') = \text{round}(M_3(d,d')),
    \nonumber\\
    M_5(d,d') &= 1 - \text{LLM}_\text{pair}(d', d),
    \quad
    M_6(d,d') = \text{round}(M_5(d',d)), \nonumber
    \\
    M_7(d,d') &= \text{sign}(\text{LLM}_\text{point}(d) - \text{LLM}_\text{point}(d')). \nonumber
\end{align}
In actuality, all component models are based on the underlying LLM (and BM25), but in our formal framework each model outputs a single prediction, thus formally we have eight components.
Interestingly, this means we are not just learning where to apply the \ac{LLM} but also how to prompt it and how to transform its predictions.

Part of our transformations are based on rounding ($M_2$, $M_4$, $M_6$, $M_7$) as previous work on \ac{PRP} found rounding to result in better ranking~\citep{qin2023rd}.
Additionally, previous work found that the order of presentation in a pairwise prompt matters, i.e., $\text{LLM}_\text{pair}(d, d') \not= 1 - \text{LLM}_\text{pair}(d', d)$~\citep{qin2023rd}.
Therefore, we also use predictions of pairs where a document is the second in the prompt ($M_5$, $M_6$).
Lastly, we found that most pointwise \ac{LLM} predictions of relevance are close to zero or one, making them uncalibrated for ranking.
As a solution, we include pseudo-pairwise predictions that are simply the sign of the difference between two pointwise predictions ($M_7$), this adds stability to the optimization without any re-calibration of our \ac{LLM}.

Our selection policy is still based on the original pointwise and pairwise predictions, we simply consider one of the transformed predictions selected if its underlying \ac{LLM} prediction(s) are selected.
Thus, our selections are still a sampled vector $s_\text{point} \in \mathds{Z}_2^{K_0}$ and a sampled matrix $s_\text{pair}\in \mathds{Z}_2^{K_0 \times K_0}$ from which the corresponding selections $(s_1,\ldots,s_7)$ can be inferred by a simple mapping:
\begin{align}
    s_\text{point}(d) = 1 &\longleftrightarrow s_1(d) = 1 \land s_2(d) = 1, \nonumber\\
    s_\text{pair}(d,d') = 1 &\longleftrightarrow s_3(d,d') = 1 \land s_4(d,d') = 1, \nonumber\\
    s_\text{pair}(d'\!,d) = 1 &\longleftrightarrow s_5(d,d') = 1 \land s_6(d,d') = 1, \nonumber\\
    s_\text{point}(d) = 1 \land s_\text{point}(d') = 1 &\longleftrightarrow s_7(d,d') = 1.
\end{align}

Similarly, our scoring function $f$ consists of eight linear subfunctions $f = (f_0, f_1, \ldots, f_7)$, each corresponding to one of the models.
The first function $f_0$ is the same as in Eq.~\ref{eq:f0}; the functions corresponding to pointwise components $(f_1, f_2)$ follow Eq.~\ref{eq:f1} but have separate $B_r$ and $C_r$ variables per function;
likewise, the functions for the pairwise components $(f_3,\ldots, f_7)$ follow Eq.~\ref{eq:f2} with separate variables per function as well.
Our final scoring function becomes:
\begin{align}
    &M^*(d \mid f, s) 
    = 
    f_0(R_0^{-1}(d))
    + f_1(s_1(d), s_1(d) \cdot M_1(d), R_0^{-1}(d))
    \nonumber \\ & \qquad\qquad\quad\;\;
    + f_2(s_2(d), s_2(d) \cdot M_2(d), R_0^{-1}(d)) \; +
    \\
    &  \sum_{i\in\{3,\ldots,7\}} \sum_{d' \in R_0} f_i(s_i(d,d'), s_i(d,d') \cdot M_i(d,d'),R_0^{-1}(d),R_0^{-1}(d')).
    \nonumber
\end{align}
Analogous to our simple model, this can be rewritten to matrix operations following the approach of Section~\ref{sec:matrixnotation}.

{
\setlength{\tabcolsep}{-0.5pt}
\renewcommand{\arraystretch}{0.0}
\begin{figure*}[t]
\centering
\vspace{-0.75\baselineskip}
    \begin{tabular}{r r r}
    \multicolumn{3}{c}{\hspace{0.52cm}\includegraphics[scale=0.35]{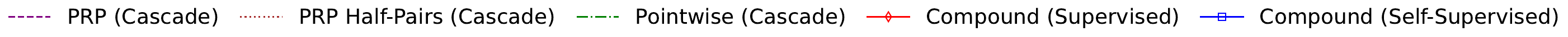}}
    \vspace{-0.1cm} \\
    \rotatebox[origin=lt]{90}{\hspace{1.1cm}\footnotesize  nDCG@1000}%
    \includegraphics[scale=0.388]{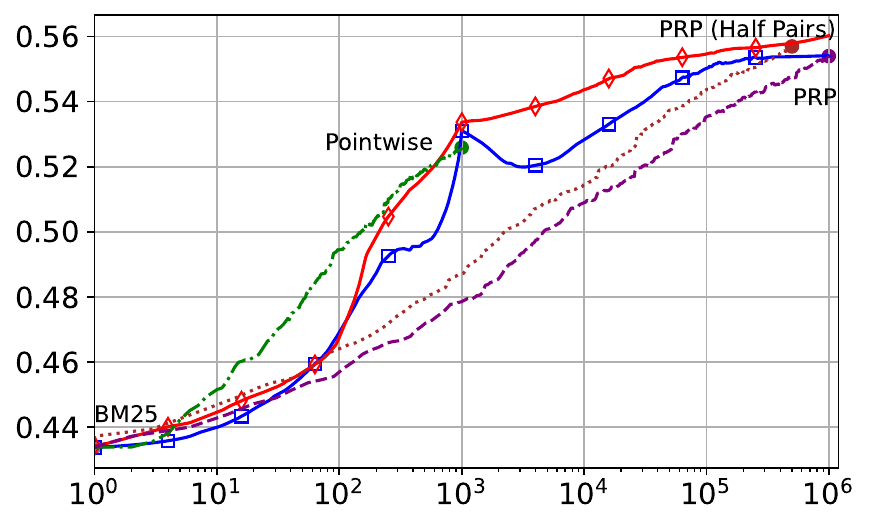}\vspace{-0.045cm}&
    \rotatebox[origin=lt]{90}{\hspace{1.1cm}\footnotesize  nDCG@100}%
    \includegraphics[scale=0.388]{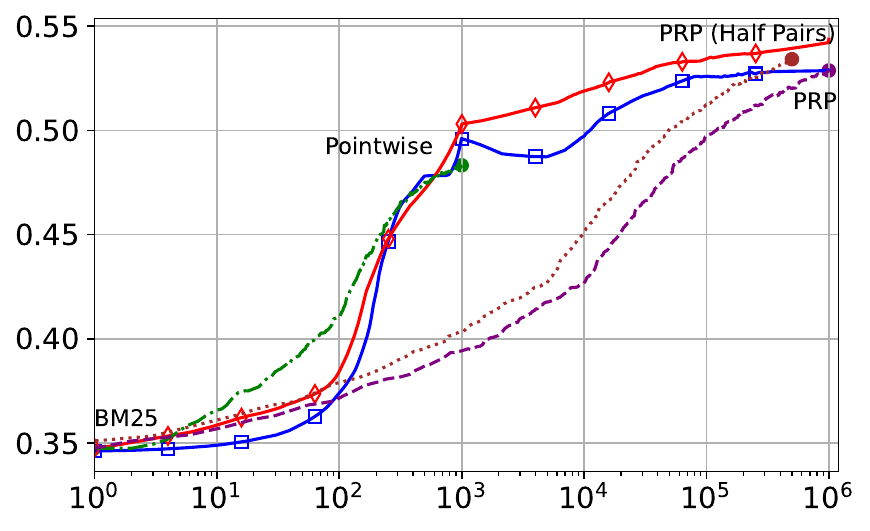}\vspace{-0.045cm}&
    \rotatebox[origin=lt]{90}{\hspace{1.1cm}\footnotesize  nDCG@25}%
    \includegraphics[scale=0.388]{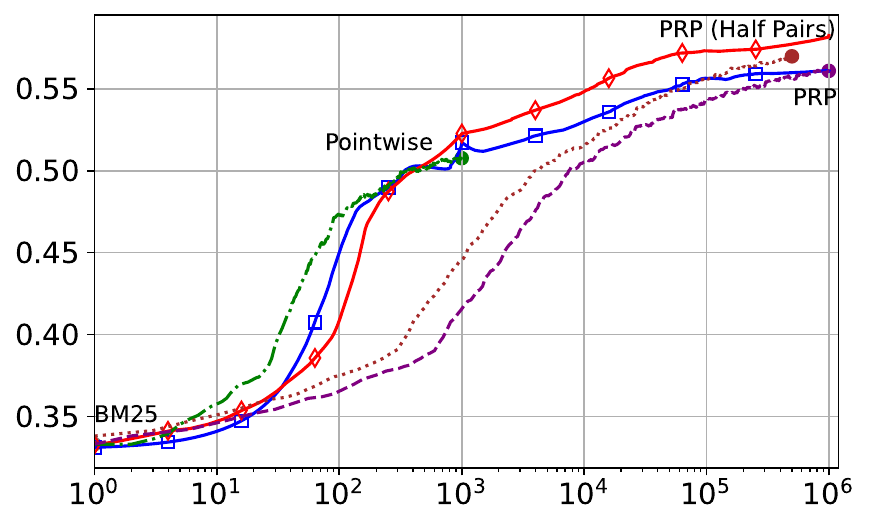}\vspace{-0.045cm}\\
    \rotatebox[origin=lt]{90}{\hspace{1cm}\footnotesize  distil-DCG@1000}%
    \includegraphics[scale=0.388]{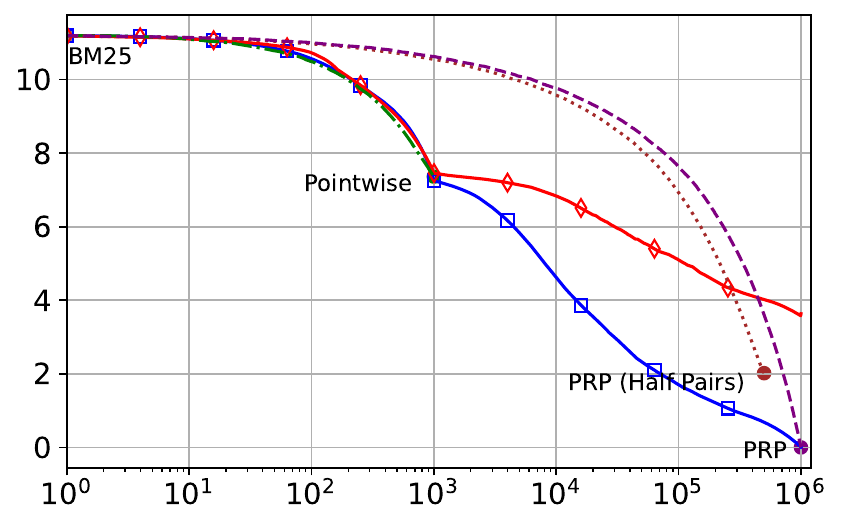} &
    \rotatebox[origin=lt]{90}{\hspace{1cm}\footnotesize  distil-DCG@100}%
    \includegraphics[scale=0.388]{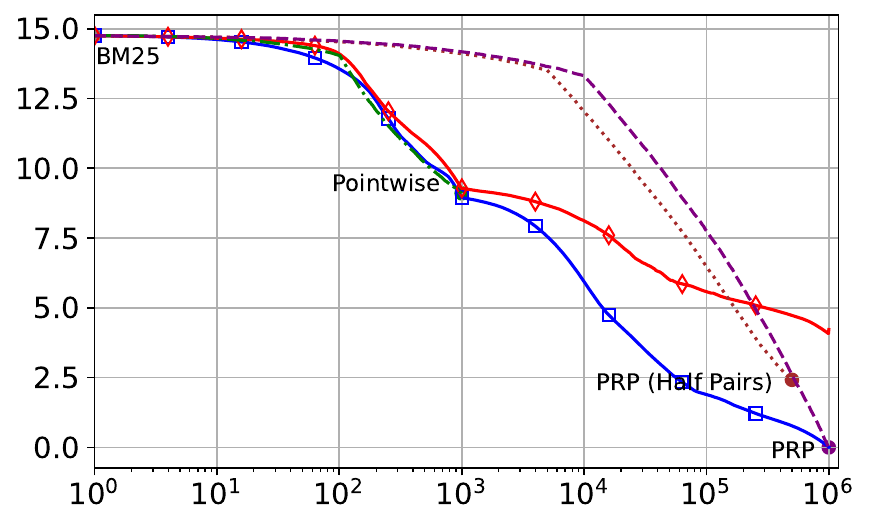} &
   \rotatebox[origin=lt]{90}{\hspace{1cm}\footnotesize  distil-DCG@25}%
    \includegraphics[scale=0.388]{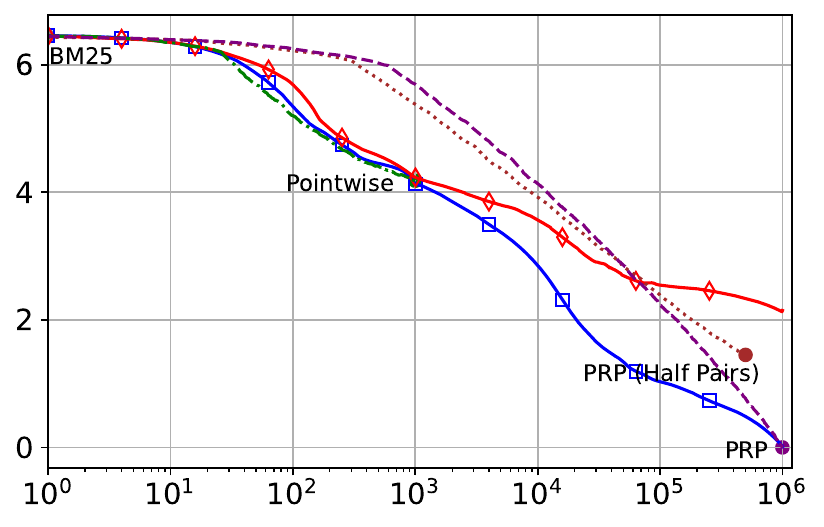} \\
    \multicolumn{1}{c}{ \footnotesize\hspace{0.44cm}Number of LLM Calls per Query ($N$)} &
    \multicolumn{1}{c}{ \footnotesize\hspace{0.44cm}Number of LLM Calls per Query ($N$)} &
    \multicolumn{1}{c}{ \footnotesize\hspace{0.44cm}Number of LLM Calls per Query ($N$)}
    \end{tabular}
    \vspace{-1\baselineskip}
    \caption{
    Effectiveness-efficiency trade-off curves (averages over 50 runs),
    created by optimizing compound systems with various trade-off weight, and varying the top-$K$ to re-rank for cascade systems.
    Annotations indicate highest baseline effectiveness.
    }
    \label{fig:curves}
    \vspace{-\baselineskip}
\end{figure*}
}

\subsection{Selection policy and scoring function}
As discussed in the final paragraph Section~\ref{sec:matrixnotation}, we use two underlying neural networks to build the variable matrices to increase regularity and generalization capability.
Both networks have three hidden layers of 64 units with sigmoid activation functions.
The first network takes as input a rank $r \in \{1, \ldots, |R_0|\}$ (corresponding to an index for a vector) and outputs a value for $A_r$; values for $B_r$ and $C_r$ for each pointwise model; and a value for the selection policy $\pi(s_\text{point}[r] = 1)$ (Eq.~\ref{eq:selectionpolicy}).
Similarly, the second network takes as input a pair of ranks $(r,r')$ and outputs a value for $A_{r,r'}$; values for $B_{r,r'}$ and $C_{r,r'}$ for each pairwise model; and a value for the selection policy $\pi(s_\text{pair}[r,r']) = 1$.
Thus, with only two small neural networks, every variable for our scoring function is generated.

\subsection{Optimization of trade-off curves}

Optimization is performed through stochastic gradient descent, our entire dataset and pre-computed LLM predictions fit in memory enabling full batch optimization.
As our loss functions, in the supervised setting, we optimize the DCG@$K$ loss using the rank and cutoff approximations from Section~\ref{sec:methodrankinglosses:cutoffs}~\citep{jagerman2022rax};
in the self-supervised setting, we apply the ranking distillation loss from Section~\ref{sec:rankingdistillation} with DCG@$K$ weights, we refer to this loss as distil-DCG@$K$.
The distillation loss is computed with respect to the PRP ranking, as this is expected to maximize effectiveness.
Thus, in our supervised setting, we use labels to optimize the DCG of our system under efficiency constraints; and in the self-supervised setting, we optimize a system to produce rankings similar to PRP but at higher efficiency.

We optimize each individual system through 15000 gradient steps with the Adamax optimizer~\citep{kingma2014adam}.
At the end of optimization, the parameters are chosen that produced the highest validation loss, subsequently, the learned selection policy is made deterministic as described in Section~\ref{sec:method:deterministic} with 250 samples.

To create our trade-offs, we use 200 values for $\alpha$ (Eq.~\ref{eq:loss}) in a geometric sequence from $1$ to $10^{-5}\!$.
A compound system is optimized for each value with a different random seed, subsequently, we discard all systems that have a higher costs and a worse validation loss than others.
From the remaining points, a curve is created by linear interpolation between the points, this is required for the computing of a mean curve and its standard deviation for our evaluation.
Separate curves are generated for $K \in \{25,100,1000\}$.

\subsection{Evaluation and baselines}
Our evaluation metrics match our optimization settings: top-$K$ normalized DCG (nDCG@$K$)~\citep{jarvelin2002cumulated} and distil-DCG@$K$.
For the cutoff $K$, we always report the performance of the method that optimized a matching $K$ (e.g., for nDCG@$100$, we report the mean for our systems that optimized DCG@$100$ or distil-DCG$@100$).
All our reported results are (test-set) averages of 50 trade-off curves created by 50 independent runs of 50 different splits of data.

As baselines, we apply the pointwise ranking model and PRP as a top-$K$ re-rankers over the top-1000 of BM25, thus simulating a cascading retrieval system with a single re-ranking stage.
Additionally, we apply a version of PRP that uses half the number of pairwise comparisons, by only selecting pairwise comparisons in one direction. %
To create trade-off curves, we vary $K$ from $1$ to $1000$;
baseline results are also averages over the same 50 data splits.

\section{Experimental Results}
\subsection{Effectiveness-efficiency trade-offs}

Figure~\ref{fig:curves} displays the effectiveness-efficiency trade-off curves of our compound retrieval systems and baseline systems, in terms of nDCG@$K$ and distil-DCG@$K$ over the number of LLM calls per query ($N$).
Table~\ref{tab:ndcg} reports nDCG@$100$ for specific $N$ values of all systems with results of significance testing.

The trade-off curves for the pointwise re-ranker and PRP reveal that PRP ultimately reaches much higher nDCG, but for equal $N$, pointwise always outperforms PRP.
The salient difference is that after using $N=1000$ predictions to re-rank the entire top-1000, the pointwise system cannot increase $N$ any further.
Interestingly, PRP with half the pair predictions is more effective than standard PRP.

Surprisingly, for all $K\!\in\!\{25,100,1000\}$, our supervised compound system reaches higher nDCG@K than the baselines even at maximal costs:
It outperforms pointwise with $N\!=\!10^3$ LLM calls, PRP with $N\!=\!10^6$ and PRP half-pair with $N\!=\!\frac{10^6}{2}$.
Thus, the compound system must have learned aggregation strategies that improve the standard behavior of its component models.
We think this is achieved because the compound model can consider a mixture of predictions when re-ranking, whereas the cascading system only uses the predictions of a single model.
Thus, for example, the compound system might penalize documents ranked low by BM25, so that they require more positive pointwise and PRP predictions to reach the front of the ranking.
This is surprising since compound retrieval systems are designed for balancing effectiveness with efficiency, and not for direct effectiveness improvements.

Furthermore, the supervised compound system clearly provides the best trade-off curves for nDCG when $N\!>\!200$, where it matches and outperforms pointwise, reaches PRP's performance with over ten times fewer LLM calls, and has similar gains over PRP half-pairs.
Table~\ref{tab:ndcg} indicates that these improvements are statistically significant for nDCG@$100$.
However, when $N\!<\!200$, the nDCG from both compound systems drops below that of pointwise, thus, it appears our optimization is less effective for DCG under such extreme efficiency constraints.
We have ruled out overfitting but could not determine the cause, we speculate our gradient approximations degrade under extreme sparsity.
The self-supervised compound system does not reach the same nDCG curve as its supervised counterpart, this is likely because it is optimized for distillation instead of ranking performance.
Nevertheless, without additional labels, it provides a considerably better nDCG trade-off than PRP. %

For distil-DCG, an outstanding trade-off curve is achieved by the self-supervised compound system; it closely matches pointwise until $N\!=\!10^3$ when it starts to considerably outperform all other systems until perfect distil-DCG is reached at $N\!=\!10^6$.
This reveals that the self-supervised compound system can provide rankings much more similar to PRP's top-$1000$ re-ranking than when using PRP as a re-ranker on a smaller top-$K$.
As expected, the supervised compound system provides a curve that performs worse and does not converge at zero.
This can be explained by the fact that it outperforms PRP in terms of nDCG, it is thus intentionally converging on a different ranking than PRP, whilst this leads to significantly higher nDCG, it naturally produces a non-zero distillation loss.

To summarize, our optimized compound retrieval systems provide significantly better overall effectiveness-efficiency trade-off curves.
In particular, supervised optimization leads to the best nDCG results, including 10$\times$ efficiency improvements over PRP, but we note that our systems falter on very sparse strategies ($N\!<\!200$).

{
\setlength{\tabcolsep}{1.75pt}
\renewcommand*{\arraystretch}{0.9}
\begin{table}[t]
    \caption{
    NDCG@$100$ for varying numbers of LLM calls.
    Results are means over 50 independent runs, brackets show standard deviations;
    $\triangle$/$\triangledown$ indicate significantly higher/lower results over the best baseline ($\rho < 0.01$, two-sided t-test).
    }
    \label{tab:ndcg}
    \vspace{-\baselineskip}
    \centering
    \begin{tabular}{l c c c c }
        \toprule
        \small \# LLM calls & \small $N=10^2$ & \small $N=10^3$ & \small $N=10^4$ & \small $N=10^5$ \\
        \midrule
        \multicolumn{5}{c}{\small\emph{Cascading Retrieval Systems}} \\
        \midrule
        \small PRP
         & \small 0.373 (\footnotesize0.0311) & \small 0.395 (\footnotesize0.0321) & \small 0.427 (\footnotesize0.0322) & \small 0.497 (\footnotesize0.0329) \\
        \small PRP (Half)
        & \small 0.379 (\footnotesize0.0320) & \small 0.404 (\footnotesize0.0319) & \small 0.451 (\footnotesize0.0328) & \small 0.513 (\footnotesize0.0326) \\
        \small Pointwise 
        & \small\bf 0.411 (\footnotesize0.0326) & \small 0.483 (\footnotesize0.0335) & - & - \\
        \midrule
        \multicolumn{5}{c}{\small\emph{Compound Retrieval Systems}} \\
        \midrule
        \small Self-super.
        & \small 0.374$^\triangledown$(\footnotesize0.0285) & \small 0.496$^\triangle$(\footnotesize0.0331) & \small 0.497$^\triangle$(\footnotesize0.0368) & \small 0.526 (\footnotesize0.0331) \\
        \small Supervised
        & \small 0.384$^\triangledown$(\footnotesize0.0323) & \small\bf 0.502$^\triangle$(\footnotesize0.0336) & \small\bf 0.519$^\triangle$(\footnotesize0.0320) & \small\bf 0.534$^\triangle$(\footnotesize0.0320) \\
        \bottomrule
    \end{tabular}
    \vspace{-1.2\baselineskip}
\end{table}
}

{
\setlength{\tabcolsep}{5pt}
\renewcommand{\arraystretch}{0.1}
\begin{figure*}[t]
    \centering
    \vspace{-0.75\baselineskip}
    \begin{tabular}{c c c c}
    \footnotesize (a) Distil-DCG@$25$ loss, $N=274$.
    &
    \footnotesize (b) Distil-DCG@$100$ loss, $N=3013$.
    &
    \footnotesize (c) DCG@$25$ loss, $N=19870$.
    &
    \footnotesize (d) DCG@$25$ loss, $N=103000$.
    \\
    \includegraphics[width=0.22\textwidth]{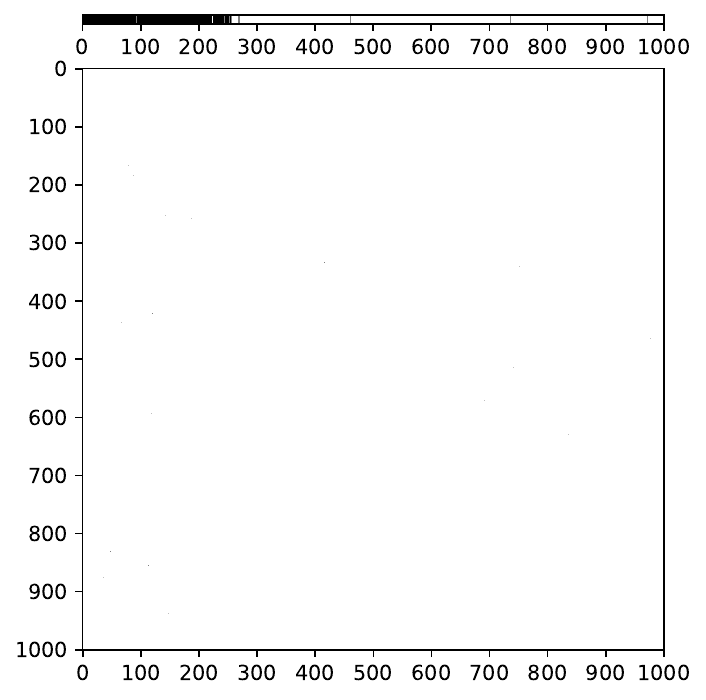}
    &
    \includegraphics[width=0.22\textwidth]{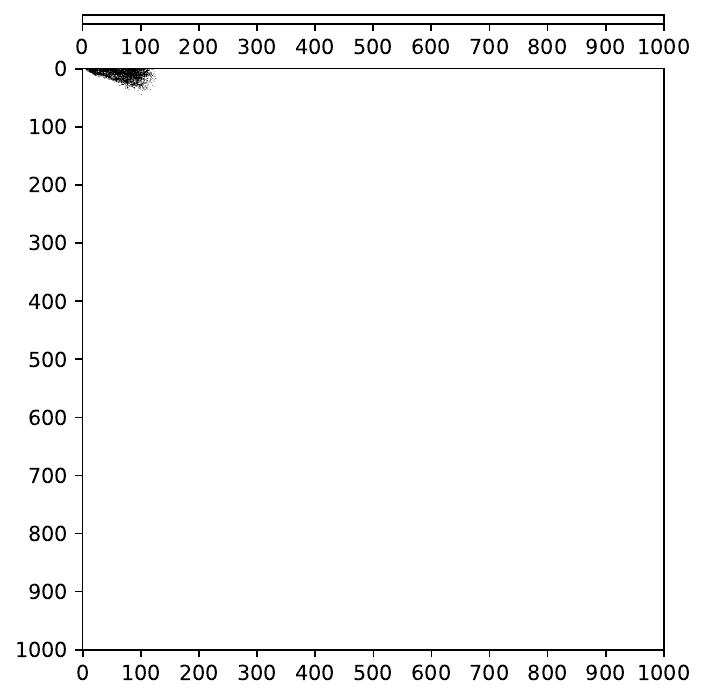}
    &
    \includegraphics[width=0.22\textwidth]{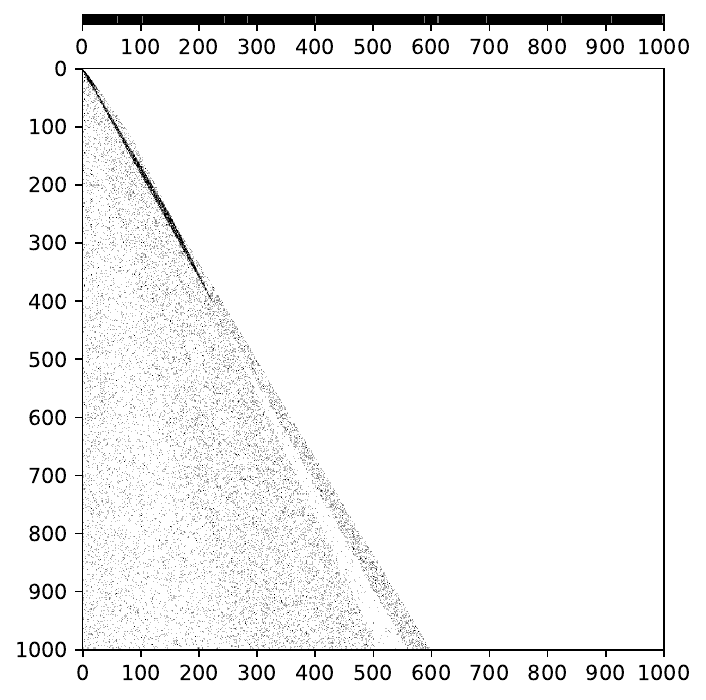}
    &
    \includegraphics[width=0.22\textwidth]{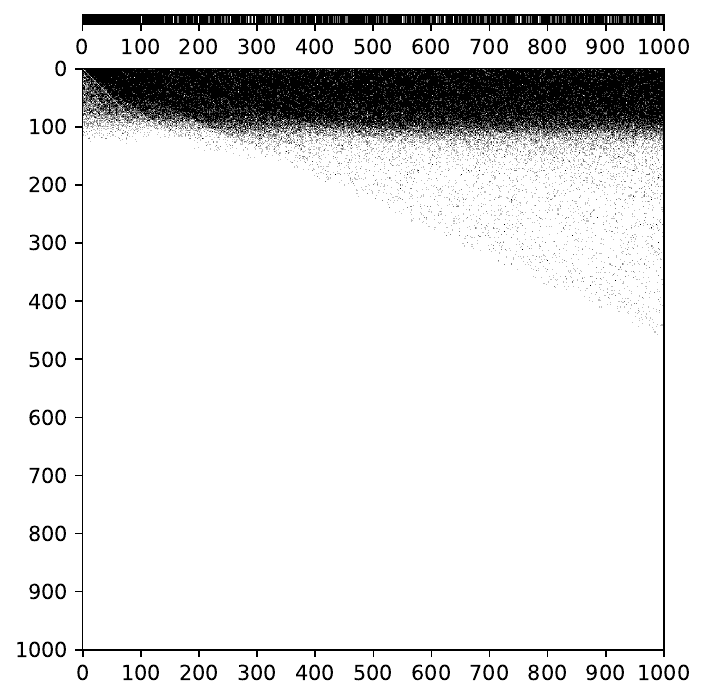}%
    \vspace{0.05cm}
    \\
    \footnotesize (e) Distil-DCG@$100$ loss, $N=103774$.
     &
    \footnotesize (f) Distil-DCG@$25$ loss, $N=212540$.
    &
    \footnotesize (g) Distil-DCG@$1000$ loss, $N=228009$.
    &
    \footnotesize (h) Distil-DCG@$100$ loss, $N=330653$.
    \\
    \includegraphics[width=0.22\textwidth]{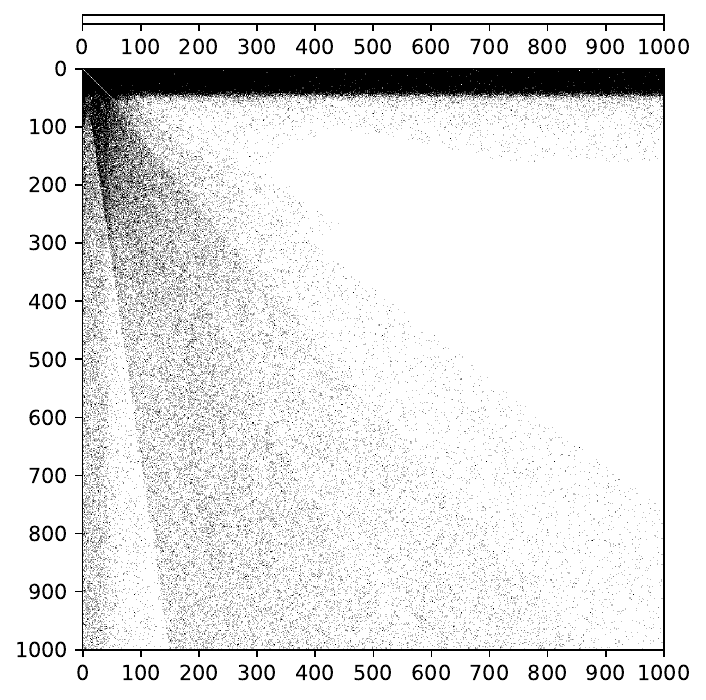}
    &
    \includegraphics[width=0.22\textwidth]{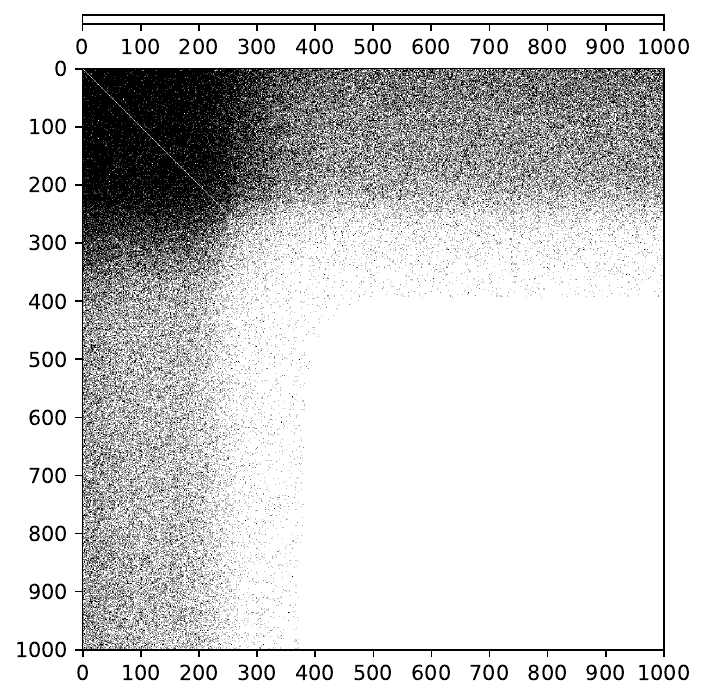}
    &
    \includegraphics[width=0.22\textwidth]{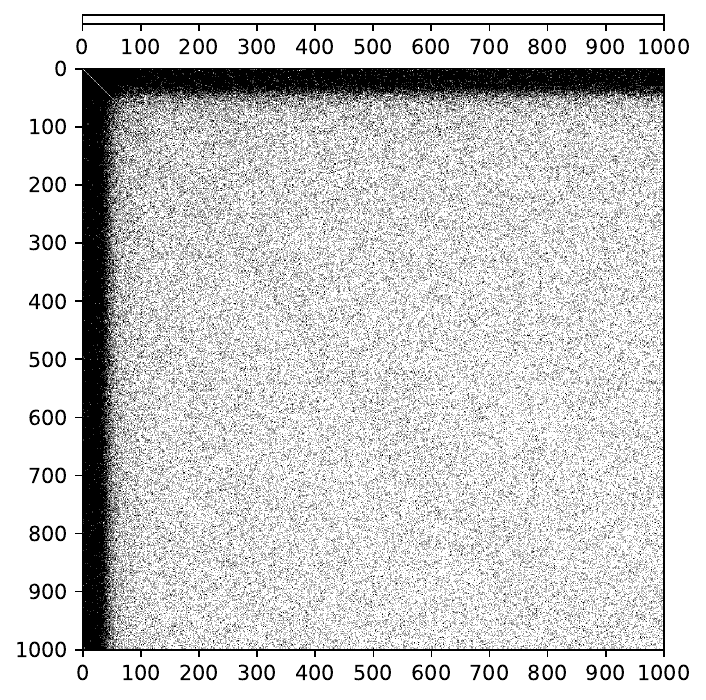}
    &
    \includegraphics[width=0.22\textwidth]{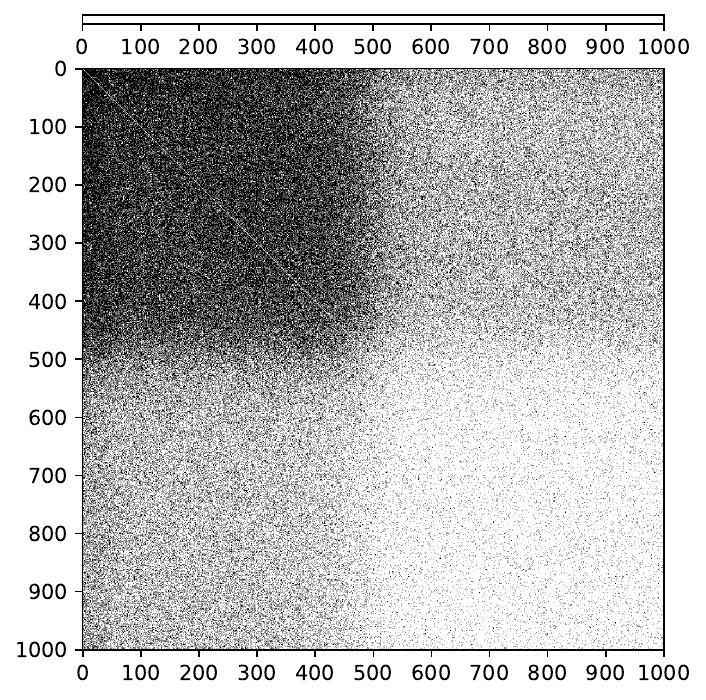}
    \\
    \end{tabular}
    \vspace{-0.9\baselineskip}
    \caption{
    Examples of (deterministic) selection policies of our optimized compound retrieval systems.
    Top bars display the selection of pointwise predictions, square matrices that of pairwise predictions (black means selected).
    Pixels indices correspond to the first-stage ranking (e.g., the $i$th pixel in the top-bar indicates whether the pointwise prediction for $i$th document in the BM25 ranking was selected).
    Above each policy is the ranking loss used for optimization and its total number of selections $N$.}
    \label{fig:policies}
    \vspace{-1.1\baselineskip}
\end{figure*}
}

\subsection{Strategies learned by compound systems.}

To investigate the behavior of our compound systems, we examined their learned selection policies.
We found these to vary widely, depending on the optimization loss and efficiency constraints applied.
Figure~\ref{fig:policies} displays eight example policies that exhibit several recurring patterns:
\begin{enumerate*}[label=\arabic*)]
    \item Example (a) learned a cascading behavior in the self-supervised setting that only gathers pointwise predictions on the top-274 of BM25.
    \item Example (b) gathers pairwise comparisons of the top-50 with the top-100 in one direction, appearing as a small triangle.
    \item Examples (c) and (d) combine pointwise and pairwise predictions, a typical behavior in the supervised setting.
    \item Many policies gather all pairwise comparison with a top-$K$ in a single direction, which appears as black bars on the matrix edges, e.g., in examples (d), (e) and (g).
    This seems to be an efficient strategy to find top documents.
    \item Example (f) and (h) display a typical pattern where all pairwise comparisons are gathered for pairs that are both in a top-$K$, fewer are gathered for pairs with only one in the top-$K$, and very few where neither are in the top-$K$.
\end{enumerate*}
Importantly, except for example (a), none of these behaviors can be recreated by a cascading retrieval system, thus, each reveals a newly discovered strategy that only compound retrieval systems can perform.

\section{Related Work}

Our work is most related to existing literature on cascading retrieval system optimization~\citep{wang2011cascade, culpepper2016dynamic, liu2022neural}.
When introducing cascading systems \citet{wang2011cascade} propose using AdaRank~\citep{xu2007adarank} to optimize the cascade structure and parameters; \citet{gallagher2019joint} extends this approach with a joint optimization of the underlying ranking models; a direction continued by \citet{qin2022rankflow} and \citet{zheng2024full}.
To the best of our knowledge, no previous work has considered alternatives to cascading retrieval system designs.

Several works use \acp{LLM} to re-rank by directly generating a permutation~\citep{sun2023chatgpt, pradeep2023rankzephyr, parry2024top, meng2024ranked}, they encode the entire document set~\citep{sun2023chatgpt, pradeep2023rankzephyr}, or predict over subsets and aggregate (a.k.a.\ listwise comparisons)~\citep{parry2024top, meng2024ranked}.
Because we find scalability essential, we only consider approaches that execute a single round of \ac{LLM} predictions in parallel.
Pointwise relevance generation~\citep{liang2022holistic} and pairwise comparisons~\citep{qin2024large} meet this requirement.
\citet{qin2024large} propose sorting strategies for \ac{PRP} with fewer pairwise predictions, but these are no longer scalable.
\citet{liusie2024efficient} also optimize a selection policy to gather pairwise comparisons from an \ac{LLM}, but for evaluating natural language generation performance, not retrieval.

\balance
\section{Conclusion and Future Work}
\label{sec:method:extensions}
\label{sec:conclusion}

This work has proposed the concept of compound retrieval systems as any retrieval system that uses multiple prediction models to create rankings.
This is a generalization of cascade retrieval systems that are limited to sequential top-$K$ ranking.
Furthermore, we have introduced a framework for optimizing compound retrieval systems for a given effectiveness-efficiency trade-off in supervised or self-supervised settings.
Our framework optimizes system design through two fundamental tasks:
\begin{enumerate*}[label=(\roman*)]
    \item learning what predictions to gather from component models, and
    \item learning how to aggregate these predictions to construct a ranking.
\end{enumerate*}
In our experiments, we optimized compound retrieval systems that combine BM25 with an LLM under pointwise and pairwise prompting.
Our experimental results show that our optimized systems provide significantly better trade-off curves than its component models (applied in a cascading system), but leave room for improvement under extreme efficiency constraints.
An analysis of the learned policies reveals that our compound systems apply a diverse set of strategies that adapt to the optimization objective, including existing strategies such as top-$K$ re-ranking, but mostly consisting of never-before-seen strategies that are unique to compound retrieval systems.
Lastly, we note that many of the observed advantages were achieved without supervision of relevance labels, indicating a very wide applicability.

This work challenged the cascading paradigm in the \ac{IR} field, and thereby, revealed an enormous potential for optimized compound retrieval systems, that we hope opens many exciting new directions for future research.
Finally, there are limitations of our work that also hint at the many possible extensions of our framework:
Firstly, the compound system we have presented in this work only performs a single re-ranking step, and accordingly, was also only compared with single re-ranking cascading systems.
A natural next direction is to extend our framework to allow for intermediate re-ranking steps, i.e., with an approach as \citet{gallagher2019joint}, this would make it a generalization of multi-stage cascades.
Secondly, our scoring aggregation functions were limited to simple linear transformations, more complex non-linear functions could potentially improve performance much further.
Thirdly, our component models were limited to pointwise and pairwise prediction models, but many other models could be included, e.g., for set-wise predictions, query-transformations, etc.
With the introduction of compound retrieval systems, the possibilities appear endless.

\begin{acks}
This research was supported by the Google Visiting Researcher program.
Any opinions, findings and recommendations expressed in this work are those of the authors and are not necessarily shared or endorsed by their respective employers or sponsors.

We thank Don Metzler, Michael Bendersky, Mohanna Hoveyda and our reviewers for useful discussions and constructive feedback.
\end{acks}

\balance
\bibliographystyle{ACM-Reference-Format}
\bibliography{references}

\end{document}